\documentclass[prd,amsmath,amssymb,twocolumn,superscriptaddress,nofootinbib]{revtex4-2}

\usepackage{graphicx}
\graphicspath{{figures/}}

\usepackage{xcolor,color}
\usepackage{url}
\usepackage{mathtools}
\usepackage{makecell}
\usepackage{dcolumn}
\usepackage{bm}
\usepackage{xspace}
\usepackage{multirow}
\usepackage{epstopdf}

\usepackage[colorlinks = true,
            linkcolor = red,
            urlcolor  = blue,
            citecolor = red,
            anchorcolor = blue]{hyperref}

\newcommand{\mus}{$\mu$s\xspace}
\newcommand{\Bi}{$^{214}$Bi\xspace}
\newcommand{\Po}{$^{214}$Po\xspace}
\newcommand{\BiPo}{$^{214}$Bi-$^{214}$Po\xspace}
\newcommand{\Rn}{$^{222}$Rn\xspace}
\newcommand{\Beta}{$\beta$\xspace}

\newcommand{\Efield}{\mathcal{E}}
\newcommand{\Wion}{W_{\text{ion}}}

\newcommand{\BetaBi}{$\beta_\text{Bi}$\xspace}
\newcommand{\AlphaPo}{$\alpha_\text{Po}$\xspace}

\begin{document}

\title{Measurement of ambient radon progeny decay rates and energy spectra in liquid argon using the MicroBooNE detector}

\newcommand{\ANL}{Argonne National Laboratory (ANL), Lemont, IL, 60439, USA}
\newcommand{\Bern}{Universit{\"a}t Bern, Bern CH-3012, Switzerland}
\newcommand{\BNL}{Brookhaven National Laboratory (BNL), Upton, NY, 11973, USA}
\newcommand{\UCSB}{University of California, Santa Barbara, CA, 93106, USA}
\newcommand{\Cambridge}{University of Cambridge, Cambridge CB3 0HE, United Kingdom}
\newcommand{\CIEMAT}{Centro de Investigaciones Energ\'{e}ticas, Medioambientales y Tecnol\'{o}gicas (CIEMAT), Madrid E-28040, Spain}
\newcommand{\Chicago}{University of Chicago, Chicago, IL, 60637, USA}
\newcommand{\Cincinnati}{University of Cincinnati, Cincinnati, OH, 45221, USA}
\newcommand{\CSU}{Colorado State University, Fort Collins, CO, 80523, USA}
\newcommand{\Columbia}{Columbia University, New York, NY, 10027, USA}
\newcommand{\Edinburgh}{University of Edinburgh, Edinburgh EH9 3FD, United Kingdom}
\newcommand{\FNAL}{Fermi National Accelerator Laboratory (FNAL), Batavia, IL 60510, USA}
\newcommand{\Granada}{Universidad de Granada, Granada E-18071, Spain}
\newcommand{\Harvard}{Harvard University, Cambridge, MA 02138, USA}
\newcommand{\IIT}{Illinois Institute of Technology (IIT), Chicago, IL 60616, USA}
\newcommand{\KSU}{Kansas State University (KSU), Manhattan, KS, 66506, USA}
\newcommand{\Lancaster}{Lancaster University, Lancaster LA1 4YW, United Kingdom}
\newcommand{\LANL}{Los Alamos National Laboratory (LANL), Los Alamos, NM, 87545, USA}
\newcommand{\Louisiana}{Louisiana State University, Baton Rouge, LA, 70803, USA}
\newcommand{\Manchester}{The University of Manchester, Manchester M13 9PL, United Kingdom}
\newcommand{\MIT}{Massachusetts Institute of Technology (MIT), Cambridge, MA, 02139, USA}
\newcommand{\Michigan}{University of Michigan, Ann Arbor, MI, 48109, USA}
\newcommand{\Minnesota}{University of Minnesota, Minneapolis, MN, 55455, USA}
\newcommand{\Nankai}{Nankai University, Nankai District, Tianjin 300071, China}
\newcommand{\NMSU}{New Mexico State University (NMSU), Las Cruces, NM, 88003, USA}
\newcommand{\Oxford}{University of Oxford, Oxford OX1 3RH, United Kingdom}
\newcommand{\Pitt}{University of Pittsburgh, Pittsburgh, PA, 15260, USA}
\newcommand{\Rutgers}{Rutgers University, Piscataway, NJ, 08854, USA}
\newcommand{\SLAC}{SLAC National Accelerator Laboratory, Menlo Park, CA, 94025, USA}
\newcommand{\SDSMT}{South Dakota School of Mines and Technology (SDSMT), Rapid City, SD, 57701, USA}
\newcommand{\Maine}{University of Southern Maine, Portland, ME, 04104, USA}
\newcommand{\Syracuse}{Syracuse University, Syracuse, NY, 13244, USA}
\newcommand{\TelAviv}{Tel Aviv University, Tel Aviv, Israel, 69978}
\newcommand{\Tennessee}{University of Tennessee, Knoxville, TN, 37996, USA}
\newcommand{\UTA}{University of Texas, Arlington, TX, 76019, USA}
\newcommand{\Tufts}{Tufts University, Medford, MA, 02155, USA}
\newcommand{\UCL}{University College London, London WC1E 6BT, United Kingdom}
\newcommand{\VTech}{Center for Neutrino Physics, Virginia Tech, Blacksburg, VA, 24061, USA}
\newcommand{\Warwick}{University of Warwick, Coventry CV4 7AL, United Kingdom}
\newcommand{\Yale}{Wright Laboratory, Department of Physics, Yale University, New Haven, CT, 06520, USA}

\affiliation{\ANL}
\affiliation{\Bern}
\affiliation{\BNL}
\affiliation{\UCSB}
\affiliation{\Cambridge}
\affiliation{\CIEMAT}
\affiliation{\Chicago}
\affiliation{\Cincinnati}
\affiliation{\CSU}
\affiliation{\Columbia}
\affiliation{\Edinburgh}
\affiliation{\FNAL}
\affiliation{\Granada}
\affiliation{\Harvard}
\affiliation{\IIT}
\affiliation{\KSU}
\affiliation{\Lancaster}
\affiliation{\LANL}
\affiliation{\Louisiana}
\affiliation{\Manchester}
\affiliation{\MIT}
\affiliation{\Michigan}
\affiliation{\Minnesota}
\affiliation{\Nankai}
\affiliation{\NMSU}
\affiliation{\Oxford}
\affiliation{\Pitt}
\affiliation{\Rutgers}
\affiliation{\SLAC}
\affiliation{\SDSMT}
\affiliation{\Maine}
\affiliation{\Syracuse}
\affiliation{\TelAviv}
\affiliation{\Tennessee}
\affiliation{\UTA}
\affiliation{\Tufts}
\affiliation{\UCL}
\affiliation{\VTech}
\affiliation{\Warwick}
\affiliation{\Yale}

\author{P.~Abratenko} \affiliation{\Tufts}
\author{O.~Alterkait} \affiliation{\Tufts}
\author{D.~Andrade~Aldana} \affiliation{\IIT}
\author{L.~Arellano} \affiliation{\Manchester}
\author{J.~Asaadi} \affiliation{\UTA}
\author{A.~Ashkenazi}\affiliation{\TelAviv}
\author{S.~Balasubramanian}\affiliation{\FNAL}
\author{B.~Baller} \affiliation{\FNAL}
\author{G.~Barr} \affiliation{\Oxford}
\author{D.~Barrow} \affiliation{\Oxford}
% ...

\collaboration{The MicroBooNE Collaboration}
\thanks{microboone\_info@fnal.gov}\noaffiliation

\date{\today}

\begin{abstract}
We report measurements of radon progeny in liquid argon within the MicroBooNE time projection chamber (LArTPC).  The presence of specific radon daughters in MicroBooNE's 85 metric tons of active liquid argon bulk is probed with newly developed charge-based low-energy reconstruction tools and analysis techniques to detect correlated $^{214}$Bi-$^{214}$Po radioactive decays. Special datasets taken during periods of active radon doping enable new demonstrations of the calorimetric capabilities of single-phase neutrino LArTPCs for $\beta$ and $\alpha$ particles with electron-equivalent energies ranging from 0.1 to 3.0~MeV. By applying $^{214}$Bi-$^{214}$Po detection algorithms to data recorded over a 46-day period, no statistically significant presence of radioactive \Bi is detected, and a limit on the activity is placed at $<0.35$~mBq/kg at the 95\% confidence level. 
This bulk \Bi radiopurity limit -- the first ever reported for a liquid argon detector incorporating liquid-phase purification -- is then further discussed in relation to the targeted upper limit of 1~mBq/kg on bulk \Rn activity for the DUNE neutrino detector.
\end{abstract}

\maketitle

\section{Introduction} \label{sec:introduction}

Liquid argon (LAr) detectors are excellent devices for performing nuclear and particle physics measurements where the deposited energy is at the MeV scale or below~\cite{Rubbia:1977zz}. The ArgoNeuT~\cite{Anderson:2012vc} and MicroBooNE~\cite{ub_det} single-phase time projection chambers (LArTPCs) have used sub-MeV detection capabilities to observe final-state neutrons from GeV-scale neutrino-nucleus interactions~\cite{argo_mev}, to set new limits on the existence of millicharged particles~\cite{argo_mcp}, and to demonstrate calibration and reconstruction techniques using MeV-scale signatures~\cite{uB_ar39,uB_mev}.  The MicroBooNE, ICARUS~\cite{ICARUS:2000ipe}, and LArIAT~\cite{lariat_detpaper} Collaborations have also measured $\mathcal{O}$(10~MeV) Michel electrons~\cite{icarus_michel,ub_michel,lariat_michels}.  
Far lower in energy, the DarkSide-50 dual-phase LArTPC and DEAP single-phase scintillation detector used $\mathcal{O}$(1--100~keV) ionization signatures from electron and nuclear recoils in argon to place new limits on dark matter~\cite{DEAP-3600:2017uua,DarkSide:2018kuk,DarkSide:2018ppu,DarkSide:2018bpj}.  
While sub-MeV scale reconstruction techniques and tools are mature for dark matter LAr experiments using dual-phase LArTPC or scintillation detector technology, similar tools are in an early stage of development for single-phase LAr neutrino detectors relying primarily on charge readout technologies~\cite{Castiglioni:2020tsu,leplar_paper,Q-Pix:2022zjm}.  

At the end of this decade, the $\approx$10~kT underground single-phase LArTPCs of the DUNE experiment will be sensitive to neutrinos produced in nearby supernovae~\cite{DUNE:2020zfm}, and may ultimately serve as a probe of solar neutrinos~\cite{dune_solar,Parsa:2022mnj}, neutrinoless double-$\beta$ decay~\cite{Mastbaum:2022rhw}, and dark sector particle interactions~\cite{Avasthi:2022tjr,leplar_paper}. Other impending or proposed future efforts also plan to realize multiton-scale LAr detectors, such as the LEGEND neutrinoless double-$\beta$ decay detector~\cite{LEGEND:2021bnm} and the DarkSide-20k and Argo dark matter detectors~\cite{DarkSide-20k:2017zyg,Akerib:2022ort}.

Many of the physics goals of these future large LAr detectors
require high radiopurities to minimize backgrounds to low-energy signals.  Radon, specifically $^{222}$Rn, is a significant source of background, as its progeny generate MeV-scale $\gamma$ rays, $\beta$ particles, and $\alpha$ particles that can produce neutrons or $\gamma$ rays in secondary interactions.  In large LAr detectors, these decay products can be generated by radon diffused throughout the LAr bulk, compromising background reduction
benefits offered by detector fiducialization. LAr and liquid xenon (LXe) detectors sensitive to low-energy signals have reduced radon contamination by implementing rigorous detector material and outgassing assay campaigns
~\cite{XENON:2017fdb, XENON:2020fbs, XENON:2021mrg,  luxzeplin_cleanliness, EXO-200:2021srn, panda_screening}.
These experiments have also installed specialized systems capable of removing radon directly from LXe through distillation~\cite{XENONnT_rnRemoval}, as well as from gaseous argon~\cite{ABE201250, DarkSide:2014llq, XENON100:2017gsw,XENON:2021mrg} and gaseous xenon~\cite{gas_xe_rn,xenon100_rn_removal}.
Using these methods, the DarkSide-50 and DEAP-3600 dark matter experiments have achieved radon levels of \mbox{$\approx2.1$~$\mu$Bq/kg}~\cite{DarkSide:2018kuk} and \mbox{$0.15$~$\mu$Bq/kg}~\cite{PhysRevD.100.022004} in their bulk LAr volumes, respectively.

Existing methods of active radio-purification may not be suitable for large next-generation experiments with LAr or LXe. Gas-phase impurity filtration technologies relying on evaporation and subsequent recondensing of the bulk LAr may not be able to achieve the throughput required for timely full-volume purification.  
In addition, as has been demonstrated for the case of electronegative impurities~\cite{Adamowski:2014daa,Andrews:2009zza}, liquid-phase argon may be less susceptible to radon contamination than the gaseous phase, indicating potential benefits in minimizing evaporation of the bulk LAr.  

Stringent radiopurity requirements for massive next-generation LAr and LXe detectors highlight the need for more dedicated liquid-phase purification research and development (R\&D).  The DUNE Collaboration aims to achieve a bulk radon contamination of \mbox{$<1$~mBq/kg} in its baseline 10 kT LArTPC modules in service to its diverse MeV-scale physics program~\cite{Avasthi:2022tjr,JuergenTalk2}. The DarkSide-20k and Argo dark matter experiments aim for \mbox{$<2$~$\mu$Bq/kg}, about three orders of magnitude lower than the nominal DUNE expectation, and in line with the purity achieved in the smaller DarkSide-50 detector~\cite{DarkSide-20k:2017zyg,Akerib:2022ort}. 

The MicroBooNE Collaboration has shown that its electronegative impurity filtration system also removed radon intended to be actively doped into its LAr bulk~\cite{ub_radon}.  
After introducing a gaseous radon source into its circulation system and bypassing the LAr filtration stage, the rate of MeV-scale signatures in the wire readout data consistent with time-correlated \BiPo decays increased. When LAr filtration was reenabled, this rate gradually returned to its steady-state baseline level measured prior to the introduction of the radon source. 
Subsequent Geiger counting surveys revealed elevated radioactivity levels in oxygen-removing filter skids containing high-area copper-impregnated aluminum pellets~\cite{ filter_ref_sieve, filter_ref_pellets}.
This unexpected demonstration refutes previous conjectures in the literature that large-throughput liquid-phase electronegative impurity filters introduce large amounts of radon into LAr detectors such as MicroBooNE~\cite{DarkSide-20k:2017zyg}. 
It also stresses the importance of studying the absolute bulk radon purity of detectors that incorporate liquid-phase filters of this type.  

In this paper, we probe the presence of radon in the MicroBooNE LArTPC by measuring the activity of specific progeny in its decay chain. This measurement expands upon the results of the earlier study demonstrating radon filtration~\cite{ub_radon}, with newly developed charge-based low-energy LArTPC reconstruction tools and a more refined analysis. New background subtraction techniques allow for accurate measurements of tagged \BiPo decays and enable the calorimetric reconstruction of their MeV-scale decay products. We use our results to set an upper limit on \Bi levels in the MicroBooNE bulk LAr of \mbox{$<0.35~\text{mBq/kg}$} at the 95\% confidence level. We then estimate the corresponding ambient \Rn activity and discuss this estimate in the context of DUNE's radiopurity requirements.

We begin with a description of the MicroBooNE detector and datasets used in Sec.~\ref{sec:microboone}.  Sections~\ref{sec:reconstruction} and~\ref{sec:analysis} then describe the MeV-scale reconstruction framework and analysis used to tag and measure \BiPo decays.  Section~\ref{sec:simulation} describes Monte Carlo (MC) simulations and data-MC comparisons used to validate reported \BiPo detection efficiencies and reconstructed energy spectra. Measured activity levels are then reported in Sec.~\ref{sec:results}, and conclusions are given in Sec.~\ref{sec:conclusion}.

\section{MicroBooNE detector and datasets} \label{sec:microboone}

MicroBooNE was a single-phase LArTPC detector located in the Booster Neutrino Beamline at Fermi National Accelerator Laboratory that operated from 2015 to 2021.  The primary component was a $2.56 \times 2.33 \times 10.37$~m$^3$ TPC containing 85~metric tons of purified LAr.  The TPC and an accompanying light collection system were contained within a cylindrical cryostat containing 170~metric tons of purified LAr. Supporting components, including readout and triggering electronics, high- and low-voltage supplies, and liquid argon filtration and monitoring systems, were inside the Liquid Argon Test Facility building housing the cryostat.  Details of the MicroBooNE detector and support systems are presented in Ref.~\cite{ub_det}.  

In the MicroBooNE LArTPC, an electric field of 274~V/cm causes ionization electrons generated by particle interactions in the active volume to drift at a rate of 1.1~mm/$\mu$s. A maximum drift time of 2.3~ms is experienced for ionization deposited near the cathode.
The drift charge arrives at an anode consisting of three planes of conducting sense wires with 3~mm pitch between wires and 3~mm spacing between planes. Inward-facing and middle ``induction'' planes each contain 2,400 wires oriented at $\pm$60$^{\circ}$ with respect to the 3,456 vertical ``collection'' plane wires.  
Induction plane wires, voltage-biased to have minimal impact on the electric field, experience bipolar currents induced by passing ionization clouds. These ionization electrons then terminate their drift on collection plane wires, generating unipolar currents. Wire signals are digitized by readout electronics with a sampling period of 500~ns per ADC time tick. 

In normal data-taking conditions, readout of the MicroBooNE detector is triggered by an external beam signal. 
For each triggered readout, 6400 samples (3.2~ms) are saved for each wire, ensuring ample time for collection of all ionization charge present inside the TPC regardless of drift distance.
Ionization charge created after the time of triggering is also collected and recorded as particle interactions continue to occur in the spatial vicinity of existing drifting electrons. 
Digitized waveforms are then filtered and processed to perform the analysis described in this paper. The residual equivalent noise charge (ENC) on wires postfiltering is around 400~$e^-$ and 300~$e^-$ for the longest wires on the induction and collection planes, respectively~\cite{ub_noise}.
While scintillation light has played a central role in prior LAr-based radiological measurements~\cite{Amaudruz:2012hr,DarkSide:2018kuk}, MicroBooNE's light collection efficiency of $\mathcal{O}$(1--10)~photoelectrons per MeV was too low to provide meaningful information for isolated MeV-scale events, so data from the light-sensitive photomultiplier tubes are not used in this analysis.

MicroBooNE's LAr purification system was designed to remove electronegative impurities, enabling the achievement of drift electron lifetimes of several tens of milliseconds during physics data-taking~\cite{ub_cal,ub_syst}.  
A mixture of recirculated liquid argon and recondensed boil-off argon gas from the cryostat ullage was fed in series through two filters at approximately 0.6~L/s~\cite{ub_radon}.
The first filter contained 4{\AA} molecular sieve material~\cite{filter_ref_sieve}, while the second contained copper-impregnated aluminum pellets~\cite{filter_ref_pellets}.  

\begin{figure}
\includegraphics[width=0.95\columnwidth]{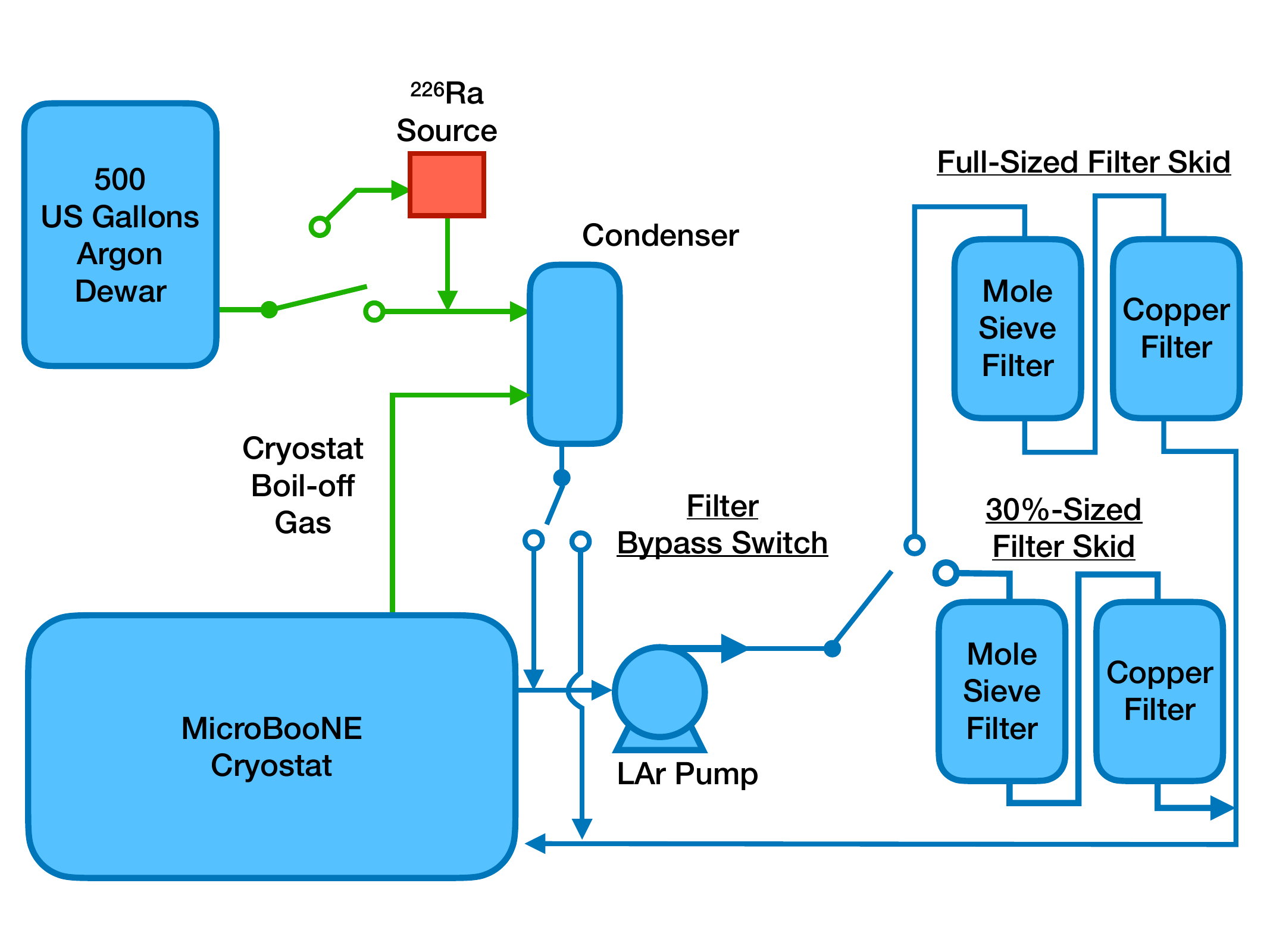}
\caption{The MicroBooNE cryogenic circulation system, including the modifications made to include a $^{226}$Ra source inline with the flow of new liquid argon for special R\&D periods described in this analysis. Gaseous and liquid argon flow is represented by green and blue lines, respectively, with arrows indicating the direction of flow. A filter bypass switch enabled a special flow configuration in which recondensed argon flowed directly into the cryostat without first passing through the filters. A second switch following the LAr pump determined whether circulating LAr flowed through the full-sized filter skid or the smaller 30\%-sized filter skid~\cite{ub_radon}. }
\label{fig:cryosystem}
\end{figure}

For a set of data-taking runs in 2021, a 500~kBq radium source ($^{226}$Ra)~\cite{radium_source} was inserted into the gas circulation line upstream from the condensers. This radium-containing argon gas was condensed and combined with recirculated argon prior to liquid filtration. Radium decays directly to $^{222}$Rn, gradually enriching the argon circulating through the system with radon. During a subset of these special runs, the recondensed $^{222}$Rn-containing LAr was routed directly into the TPC, bypassing the recirculating LAr entering the filtration system (``filter bypass'' radon doping data).  Figure~\ref{fig:cryosystem} shows a visual schematic of this special run configuration, with a more detailed description given in Ref.~\cite{ub_radon}.   The filter bypass data were used to validate the MC-reported capability of MeV-scale analysis tools by identifying and reconstructing correlated \BiPo decays in MicroBooNE.  From this dataset, \mbox{$\approx$~81,000}~events recorded over two~days using the standard filtration/circulation configurations and \mbox{$\approx~$76,000}~events recorded over two days using the filter bypass configuration were used. Due to the lack of filtration of LAr entering the cryostat, the concentration of impurities rose dramatically during this period, reducing the drift electron lifetime.

To more precisely measure the rate of \BiPo decays in liquid-filtered LAr, data from a 46-day period during a MicroBooNE physics data-taking campaign were used, recorded between June 9 and July 24, 2018. 
The \mbox{$\approx$~654,000}~detector readouts used for the analysis, representing a cumulative recorded exposure of about 35~minutes, were collected during periods when the BNB beam was not delivering neutrinos to MicroBooNE (``beam-external'' data). Instead, each readout was triggered by a low-frequency pulse delivered to the trigger system by a function generator (``unbiased'' beam-external data).  
This dataset is used to estimate cosmic backgrounds  in MicroBooNE's beam neutrino physics analyses.

\section{Low-energy reconstruction} \label{sec:reconstruction}

Here we review the novel reconstruction of MeV-scale features using MicroBooNE's charge collection system. These newly developed techniques utilize lowered thresholds to enhance sensitivity to the low-energy signals sought in this analysis.
Data processing is carried out in \texttt{LArSoft}, a common software framework used for all Fermilab LArTPCs \cite{larsoft}.

\subsection{Geometric reconstruction}
\label{sec:reconstruction_geo}

\begin{figure}
\includegraphics[width=0.95\columnwidth]{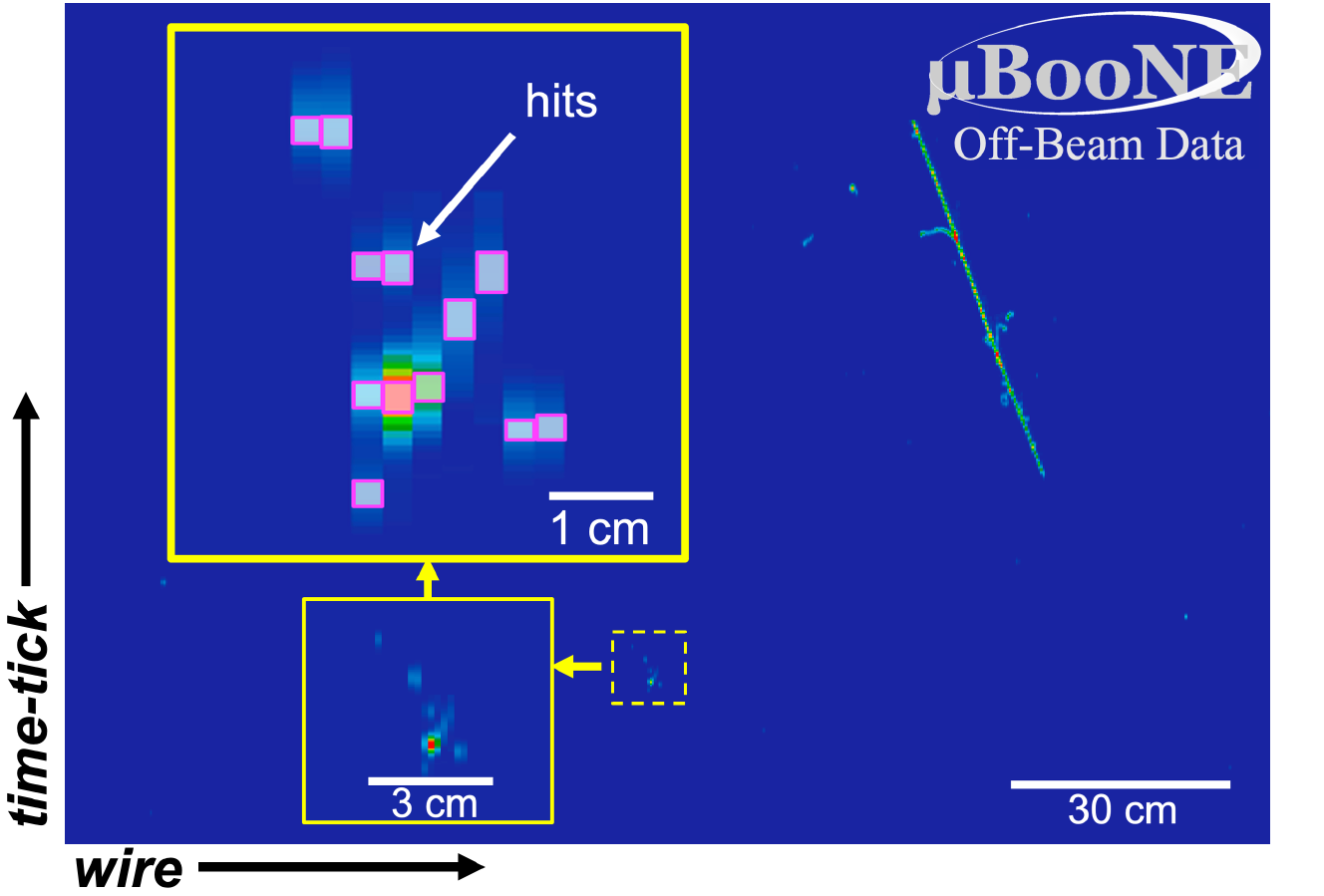}
\caption{A section of a MicroBooNE LArTPC event display from data. Each vertical column of pixels represents a single wire. The rainbow color spectrum denotes charge collected per 500~ns ADC time tick, with redder colors indicating higher charge densities. A grouping of hits across 10 wires from a randomly-selected area of interest spanning about 3~cm is represented in the zoomed inset as a set of pink boxes, where the extent of each box along the vertical (time) axis conveys the hit's RMS width. A cosmic muon track can be seen to the right.}
\label{fig:evd}
\end{figure}

Filtering and deconvolution algorithms are first applied to each digitized TPC waveform to suppress noise and account for the expected signal shape from the readout electronics.
The end result of this deconvolution process for each readout channel, visualized in Fig.~\ref{fig:evd}, is a series of charge pulses corresponding to groups of drifted electrons sensed over 3.2~ms of readout time~\cite{ub_sigproc_pt1, ub_sigproc_pt2, ub_noise}. 
An algorithm scans selected regions and fits each pulse to a Gaussian function, creating reconstructed ``hits."   Properties like amplitude, mean time, and RMS width for each hit are extracted directly from the fit~\cite{Baller_2017}. 
Plane-specific timing offsets are applied to account for the time it takes ionization electrons to drift across each 3~mm gap between adjacent readout planes. 
The pattern-recognition algorithm Pandora \cite{ub_pandora} evaluates relative orientation of reconstructed hits from each of the wire planes and identifies contiguous linelike patterns. Features that correlate across multiple planes are reconstructed by Pandora into 3D particle tracks.

Unlike tracks, MeV-scale activity creates charge depositions spanning only a few wires. To reconstruct these features, we first exclude all wire hits associated with 3D tracks longer than 5~cm. Remaining same- or adjacent-wire hits are grouped into clusters based on their relative proximity in time, with a maximum allowable separation that scales with their RMS widths. Each cluster's overall charge-weighted mean time and RMS are computed. Finally, the Gaussian integral of each hit is added up and converted from ADC counts to electrons using a plane-specific electronics calibration scale factor.  

For each hit cluster on the collection plane wires, we search for matching candidate clusters on the two induction planes. Only matches between intersecting wires are considered. If at least one matching induction plane cluster is found, a 3D  ``blip'' is reconstructed. To achieve the best reconstruction efficiency, a minimum of only two matched planes is required to form a blip for the analysis presented in this paper, though three-plane matches are common and significantly less likely to be induced by noise.

Several criteria are evaluated to determine if potential matched clusters coincide in time. The fractional overlap of the clusters' time spans must exceed 50\%. The clusters' start or end times must also coincide to within 1~\mus (2 time ticks). Finally, the clusters' charge-weighted mean times must differ by less than 80\% of the quadrature sum of the clusters' RMS values.

\begin{figure}
    \includegraphics[width=0.95\columnwidth]{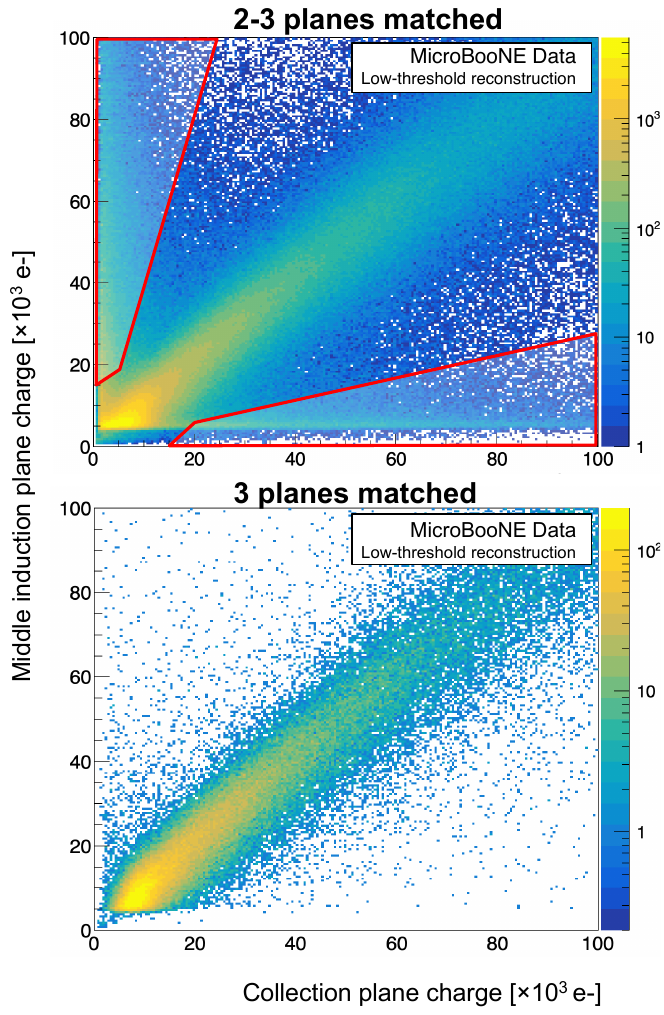}
    \caption{Distribution of integrated hit cluster charge on the collection plane and the middle induction plane for potential match candidates, for cases requiring matches in only 2--3 planes (top) and all 3 planes (bottom). Matches falling within the red regions in the top plot are rejected.}
    \label{fig:reco_qmatch}
\end{figure}

The relative integrated charge of the candidate clusters is evaluated to reject false matches. This is illustrated in Fig.~\ref{fig:reco_qmatch}, which shows the relation between charge values on the collection plane and one of the induction planes for cluster pairs satisfying the time-based criteria described above. 
Many matches are found with large charge discrepancies, where a cluster with relatively high charge on one plane is matched with a low-charge cluster on the other. When a subset of blips is selected that produce a match on all three planes, this population of charge-disparate matches disappears, suggesting that false hits are induced by electronics noise. To reject these false matches, for clusters with absolute charge differences $>$~10,000 electrons, we require the ratio of the larger cluster to the smaller cluster be \mbox{$\leq4$}.

The geometric coordinate system in the MicroBooNE volume is defined such that $\hat{x}$  is parallel to the electron drift direction, spanning the 2.56~m distance between the wire planes and the cathode. The $\hat{y}$ and $\hat{z}$ directions relate to positions along the detector's height (2.32~m) and length (10.36~m), respectively.
Each blip's \emph{y} and \emph{z} coordinates are defined by the common point of intersection of the central-most wires in the plane-matched clusters. For reconstruction of the \emph{x} coordinate, the true interaction time ($t_0$) of the particle producing the blip  must be assumed in order to convert the raw time along the wire readout signal to a physical drift time, which is then multiplied by the ionization drift velocity in the LAr volume.
In LArTPCs, $t_0$ is usually determined by an external beam signal and/or a flash of scintillation light detected by a photon detection system.
For nonbeam physics, scintillation light alone must be used for tagging $t_0$, requiring the matching of a light flash to features in the charge readout.  
In MicroBooNE, most radiological decays do not produce enough light for flash-matching. This lack of $t_0$ tagging means the \emph{x} coordinate assignment is ambiguous for the analysis presented in this paper, and is therefore not used.

\begin{figure}[t]
\includegraphics[width=0.95\columnwidth]{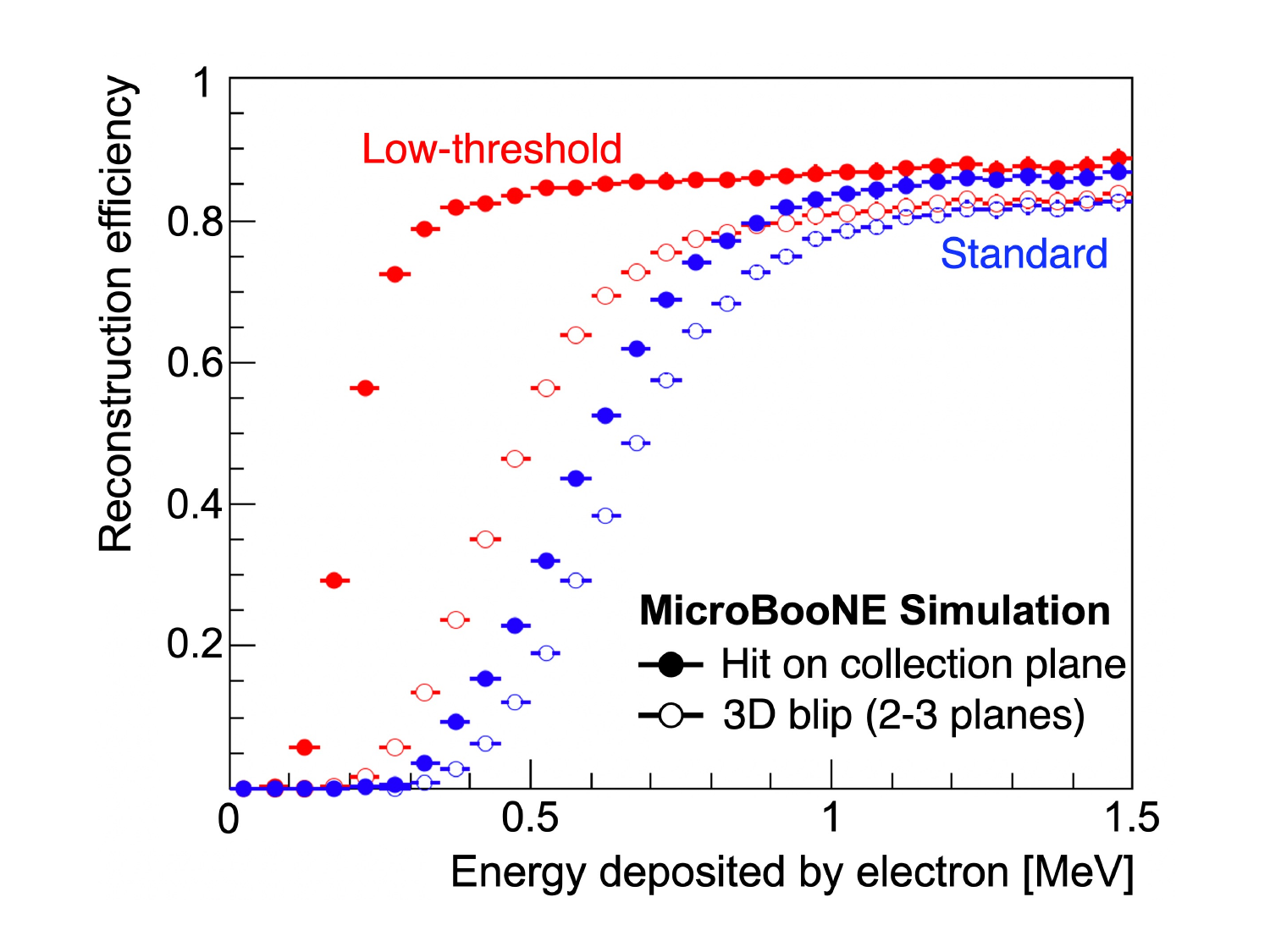}
\caption{Reconstruction efficiency as a function of energy deposited by electrons in MC simulations. Both standard MicroBooNE settings (blue) and low-threshold settings (red) are shown. The solid markers represent hit-finding performance for wire signals on the collection plane, while the unfilled markers represent 3D blips that were plane matched on at least two planes. 
The presence of nonfunctional wires on each plane limits the maximum achievable efficiencies.}
\label{fig:reco_efficiency}
\end{figure}

A simulation of the MicroBooNE detector, described in greater detail and validated using MeV-scale electrons from $^{214}$Bi decays in Sec.~\ref{sec:simulation}, is used to characterize  reconstruction performance.
Samples of low-energy electrons distributed uniformly throughout the LArTPC active volume are simulated to measure blip reconstruction efficiency. 
The efficiency is influenced by settings related to the formation of ``regions of interest'' (ROIs) in the raw signal deconvolution, and by the absolute ADC signal threshold used in the hit-finding algorithm.
Figure~\ref{fig:reco_efficiency} shows the efficiency as a function of electron-deposited energy for MicroBooNE's standard reconstruction configuration and for a special ``low-threshold'' configuration (first used in Ref.~\cite{ub_radon}) where the deconvolution ROI and hit-finding thresholds were lowered.
Unresponsive or nonfunctional wires on each plane limit the maximum achievable efficiency to $\approx85\%$ and $\approx95\%$ for the two induction planes, and $\approx90\%$ for the collection plane.
This effect is compounded for 3D plane matching, which is limited to $\approx89\%$ for matches across 2--3 planes (collection + at least one induction) and $\approx73\%$ for 3-plane matches.

\begin{table}[t]
    \centering
    \begin{tabular}{l c c}
    \hline \hline
    & \multicolumn{2}{c}{\bf 50\% Eff. Threshold [keV]} \\
        {Configuration }                  
        & \makecell[c]{Standard\\settings}
        & \makecell[c]{Low-threshold\\settings} \\ 
    \hline
        First induction plane         
            & 730               
            & 530     \\
        Second induction plane         
            & 750               
            & 540    \\
        Collection plane            
            & 620               
            & 210     \\
        3D-matched blip, 2--3 planes 
            & 670               
            & 450 \\
        3D-matched blip, 3 planes   
            & 770               
            & 600  \\
        \hline \hline
    \end{tabular}
    \caption{True electron-deposited energy at which the reconstruction efficiency reaches 50\% of its maximum achievable value for the standard and low-threshold settings.}
    \label{tab:reco_efficiency}
\end{table}

Table~\ref{tab:reco_efficiency} shows the energies at which the rising edge of the efficiency curves for these two configurations reach 50\% of the maximum achievable efficiency after accounting for nonfunctional wires. The criteria needed for forming ROIs during signal deconvolution loosened significantly in the low-threshold reconstruction, particularly for the collection plane. The result is enhanced sensitivity to lower-energy deposits on the collection plane, coupled with smaller improvements in the two induction planes. Further lowering thresholds leads to an increase in noise-induced hits being reconstructed on each plane. This not only reduces the ability to find nonambiguous matches for hits between planes, but also impacts the reconstruction of tracks. 

\subsection{Energy reconstruction} \label{sec:reconstruction_energy}

Visible energy is reconstructed using charge from the collection plane.
If $t_0$ is known, the collected charge is scaled up to account for the electrons absorbed by electronegative impurities during the drift. This correction uses the calibrated lifetime of drifting electrons, $\tau_e$, found from anode-to-cathode piercing cosmic muon tracks~\cite{ub_cal,ub_syst}. For reconstruction of ambient radiological signals presented in this analysis, the $\tau_e$ correction is not applied. 
In standard MicroBooNE operating conditions, $\tau_e$ is effectively infinite, as measured charge attenuation across the drift volume is negligible. Corrections based on each 3D blip's $y$ and $z$ coordinate are applied to account for known nonuniformities in charge collection across the collection plane~\cite{ub_cal}.

A significant fraction of ionization electrons recombine with Ar$_2^{+}$ before they drift to the wire planes. This effect must be accounted for to reconstruct the total charge deposited by a particle. The probability $\mathcal{R}$ of an electron surviving recombination 
depends on the local density of electrons, $dQ/dx$, and the electric field, $\mathcal{E}$. The energy can therefore be reconstructed using
\begin{equation} \label{eq:reco_energy}
    E_\text{reco} = \frac{Q}{ \mathcal{R}(dQ/dx,\mathcal{E}_\text{local})} \times W_\text{ion},
\end{equation}
where $Q$ is the reconstructed charge in units of electrons, and $\Wion = 23.6$~eV~\cite{PhysRevA.9.1438} is the mean energy required to produce an electron-ion pair in LAr. 

While determining $dQ/dx$ along tracks is straightforward, it is nearly impossible for MeV-scale depositions, since $dx$ cannot be reliably measured when the collected charge is concentrated on only a few readout channels~\cite{lariat_michels}.
Calorimetry at the MeV-scale is further complicated by accumulated space charge effects~\cite{Abratenko_2020} that modify the local electric field, and since electronic stopping power for electrons (and therefore recombination) increases substantially and nonlinearly for kinetic energy $\lesssim1$~MeV~\cite{argo_mev,csda}. 

\begin{figure}
\includegraphics[width=0.95\columnwidth]{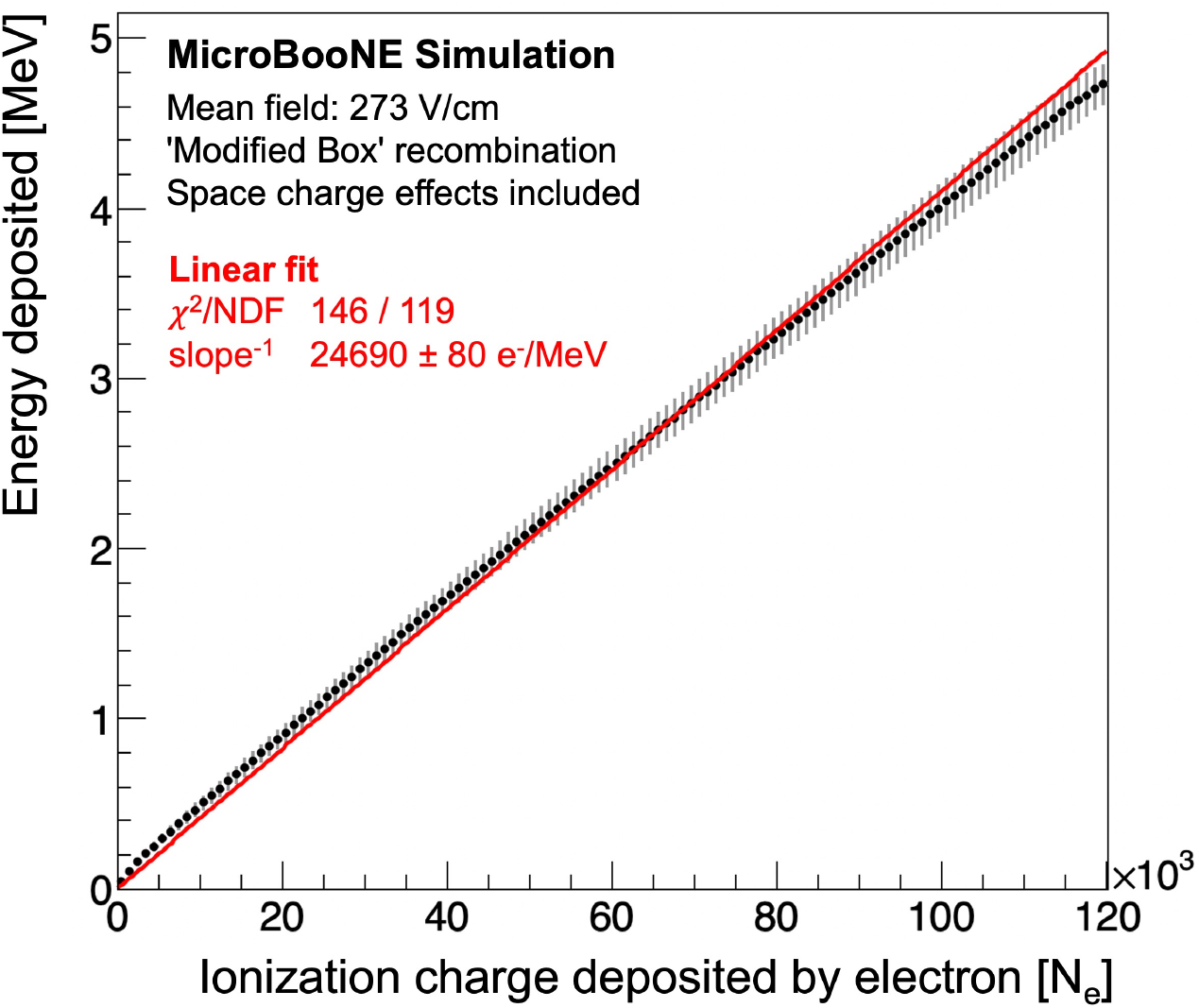}
\caption{Ionization charge as a function of corresponding energy deposition for electrons distributed uniformly in the MicroBooNE active volume. Error bars show the average variations due to nonuniformities in the electric field from accumulated space charge.}
\label{fig:charge_vs_energy}
\end{figure}

Simplifying assumptions are therefore applied to Eq.~\ref{eq:reco_energy}.
Figure~\ref{fig:charge_vs_energy} shows the relationship between the deposited energy and free ionization charge for the sample of simulated low-energy electrons described previously. The modified box model~\cite{argo_modbox} is used to calculate recombination using the local electric field. The error bars give a sense for the mean deviations caused by nonuniformity of the field due to space charge effects.  
Despite a small deviation from linearity at low energies, the relationship is approximately linear overall, with an average charge yield of $\approx$~24,700 electrons per MeV of deposited energy.
At MicroBooNE's nominal electric field, $\mathcal{E}$ = 274~V/cm, this corresponds to an equivalent electron recombination survival fraction of $\mathcal{R}~\approx0.584$, and a mean stopping power of $\langle dE/dx \rangle \approx 2.8$~MeV/cm, consistent with values calculated from the the NIST table of electronic stopping for electrons below a few~MeV~\cite{csda}. Eq.~\ref{eq:reco_energy} thus simplifies to an `electron-equivalent' energy,
\begin{equation} \label{eq:electronEquivEnergy}
     E_\text{reco} \text{ [MeVee]} = \frac{Q}{0.584} \times W_\text{ion}.
\end{equation}
For electron energy deposits between about 1.5~MeV and 3.5~MeV, this linearized reconstruction yields an energy scale bias within the range of intrinsic variations from $\Efield$-field nonuniformities. For energies in the range of 0.1--1~MeV, the energy bias ranges between about 10\% and 20\%. 
The result presented in this paper is not particularly sensitive to this reconstructed energy scale bias since it is well-understood and accurately replicated in the MC reconstruction.

\begin{figure}
\includegraphics[width=0.95\columnwidth]{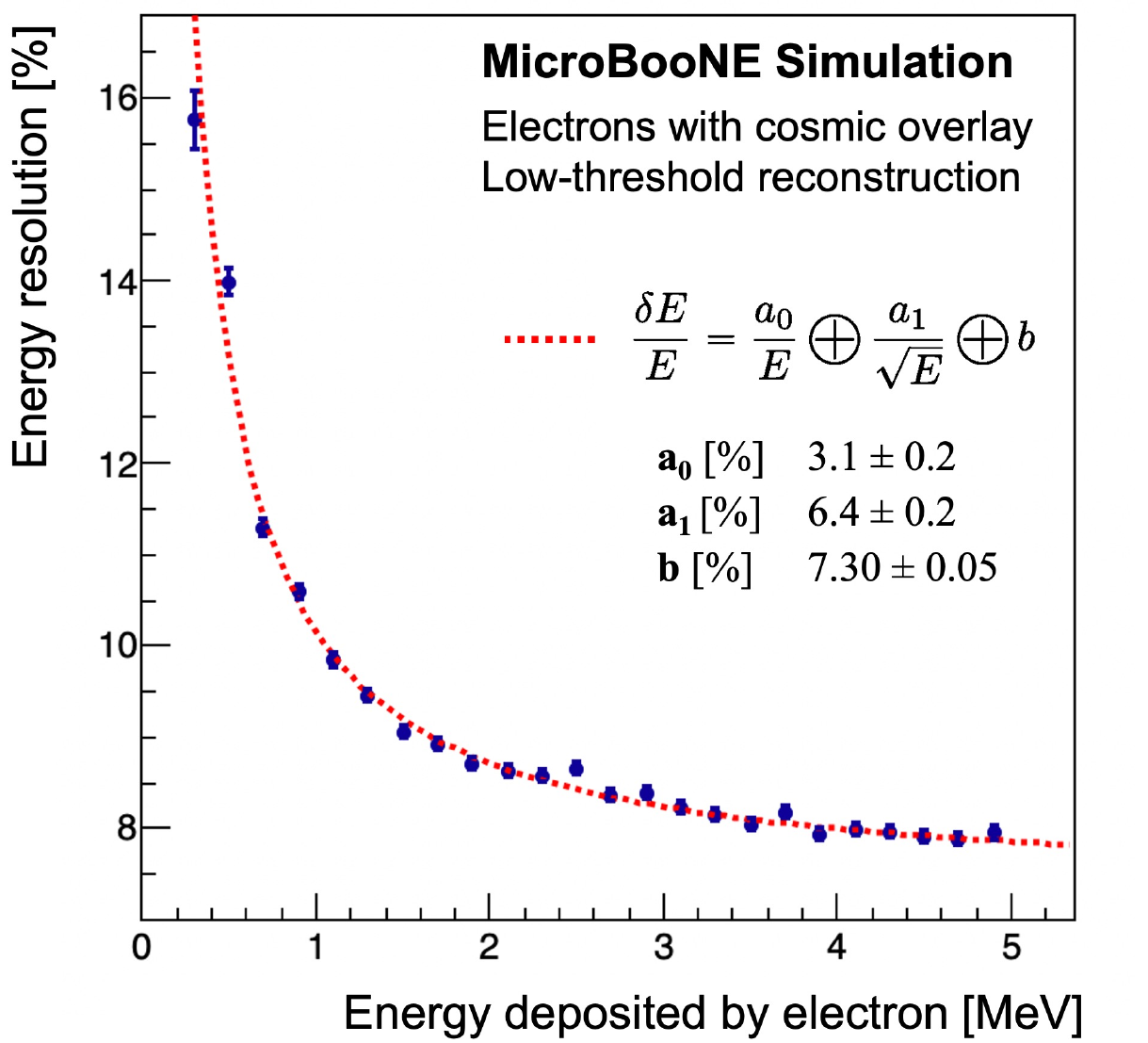}
\caption{Energy resolution from simulated electrons in the MicroBooNE TPC using low-threshold reconstruction settings. For the fit, defined in Eq.~\ref{eq:resolution}, deposited energy $E$ is in units of MeV.}
\label{fig:energy_resolution}
\end{figure}

Using this linear conversion of reconstructed charge into energy, the  energy resolution according to the MC simulation is presented in Fig.~\ref{fig:energy_resolution}. The resolution and its error-bar in each bin of deposited energy ($E_\text{dep}$) is evaluated by taking the fitted Gaussian width of the distribution of $\delta E = (E_\text{reco} - E_\text{dep})/E_\text{dep}$. 
A function that parameterizes the resolution of calorimetric detectors~\cite{kolanoski_book} is fit to the plotted MC results,
\begin{equation} 
    \label{eq:resolution}
    \frac{\delta E}{E} = \frac{a_0}{E\text{ [MeV]}} \bigoplus \frac{a_1}{\sqrt{E\text{ [MeV]}}} \bigoplus b.
\end{equation}
The terms in this function represent contributions from electronic noise ($a_0 = 3.1\%$), counting statistics ($a_1 = 6.4\%$), and reconstruction-related systematic effects ($b = 7.30\%$). This best fit corresponds to an electron energy resolution of 10\% at 1~MeV and 8\% at 5~MeV. This is well below the 10-20\% resolution needed in the DUNE detector for studying supernova neutrinos~\cite{DUNE:2020zfm}, and roughly consistent with the resolution (7\% for electrons over 5~MeV) needed for DUNE to study solar neutrinos~\cite{dune_solar}.

\section{Analysis procedure}  \label{sec:analysis}
{

\subsection{Bi-Po decay topology} \label{sec:analysis-topology}

The presence of \Rn in the TPC can be inferred by detecting decays of its progeny and correcting their measured rates to account for efficiency losses related to the plating-out of isotopes onto surfaces (as described later in Sec.~\ref{sec:results}).
A technique to identify \Bi decaying to \Po was successfully employed in the recent demonstration of filtration of radon by MicroBooNE's liquid argon purification system~\cite{ub_radon}.
The isotope \Bi ($Q_{\beta}=3.27$~MeV) decays with a half-life of 19.7~minutes, emitting an electron (or ``$\beta$ particle'') with an energy spectrum extending to the decay endpoint. The daughter \Po then decays from the same point in the TPC with a half-life of 164.3~$\mu$s~\cite{TOTH1977437}, emitting a monoenergetic 7.7~MeV $\alpha$ particle. 
Due to the electron drift velocity of 1.1~mm/\mus~\cite{ub_efield}, the temporally separated \BetaBi and \AlphaPo emissions manifest as two spatially separated signals occurring on the same readout wire(s) with an average apparent separation of 18~cm. Since the densely ionizing $\alpha$ signal is highly quenched in LAr due to  recombination and other effects~\cite{PhysRevA.35.3956,DarkSide:2016ddo}, its charge signal appears much fainter than that of the \BetaBi,  depositing only a few thousand electrons compared to the $\beta_\text{Bi}$ which deposits on average $\approx$~15,000 electrons and a maximum of $\approx$~80,000 electrons. 

The $\beta$ decay of the \Bi can also produce several low-energy $\gamma$ rays~\cite{nndc_chart} which interact primarily via Compton scattering in the surrounding LAr, creating additional blips in the vicinity of the $\beta_\text{Bi}$ signal. Since the radiation length at these energies is $\mathcal{O}$(10~cm), these displaced $\gamma$-induced blips can be mistaken for the \AlphaPo signal if they occur on the same readout channel.  
This also implies that any other $\beta$-decaying radioisotope that emits $\gamma$ rays can mimic the \BiPo signal, such as $^{214}$Pb ($Q_{\beta}=1.02$~MeV) in the \Rn decay chain. Figure~\ref{fig:evd_bipo} illustrates the \BiPo topology as it appears in a MicroBooNE event, including several potential $\gamma$ signals near the candidate $\beta_\text{Bi}$ deposition.

\begin{figure}
\includegraphics[width=0.92\columnwidth]{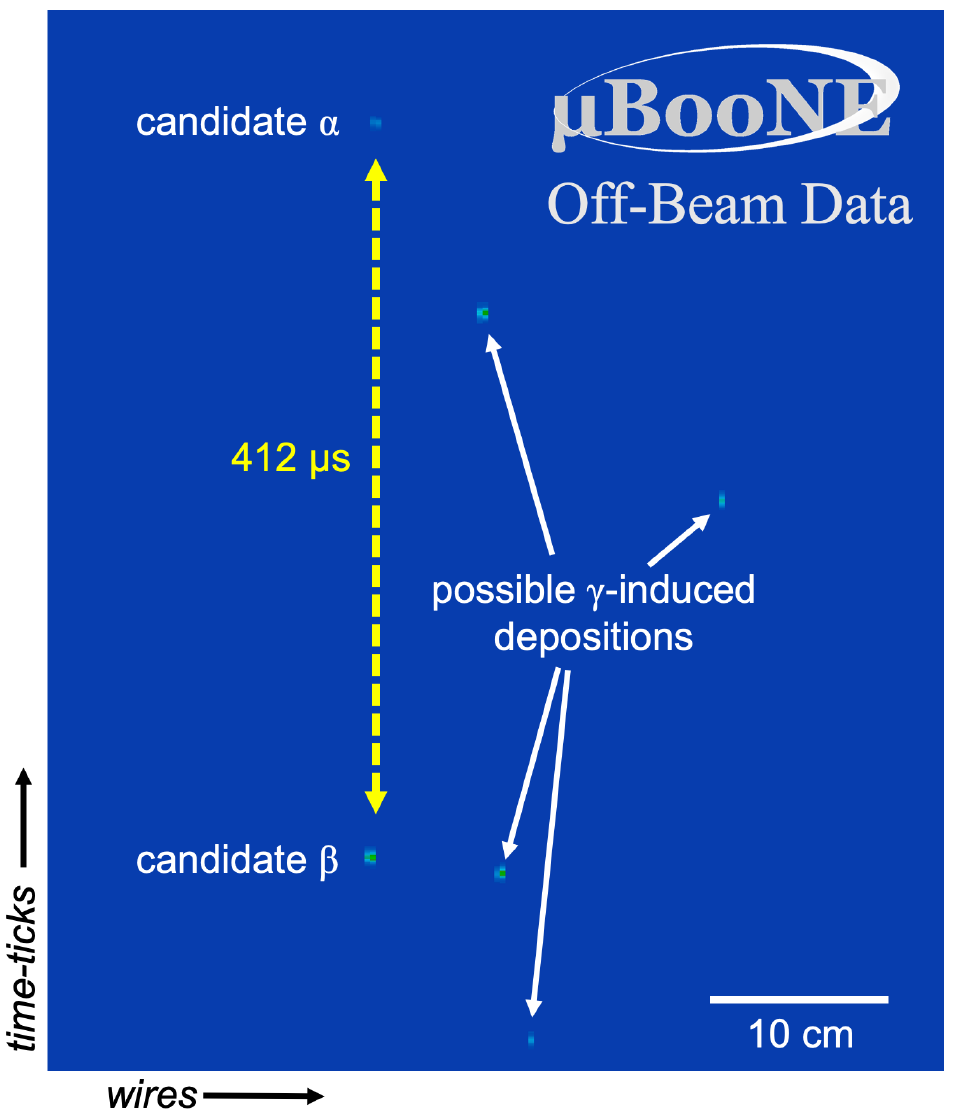}
\caption{A \BiPo decay signal candidate in an event display, including backgrounds from potential de-excitation $\gamma$ rays emitted following the decay.}
\label{fig:evd_bipo}
\end{figure}

\subsection{Signal selection}
\label{sec:analysis_sigselection}

Here we outline the selection of \BiPo (`BiPo') decay candidates. As described in Sec.~\ref{sec:microboone}, we use data collected in 2021 when a $^{226}$Ra source was used to introduce $^{222}$Rn into the MicroBooNE TPC. Our procedure is similar to that used in the study that demonstrated the removal of radon by the filtration system~\cite{ub_radon}, with modifications to improve the signal-to-background ratio.

To maximize sensitivity at lower energies, TPC data are reconstructed using the low-threshold configuration described in Sec.~\ref{sec:reconstruction}. To avoid low-energy activity induced by cosmic ray muons passing through the detector, such as $\delta$ rays, we veto all hits within 15~cm of tracks resembling through-going cosmic muons. This proximity is evaluated per-plane, in a 2D space in which each hit's drift time and wire number are converted into distance-equivalent coordinates. Remaining hits are clustered, plane-matched, and reconstructed into 3D~blips. 
Readout channels that are identified by the upstream signal deconvolution algorithm as particularly noisy are excluded from consideration. Additional requirements are enforced to reject hit clusters that are not sufficiently isolated, as well as those coinciding in time with other hits across nearby wires, a topology consistent with coherent noise.
To ensure none of the deposited energy is missed, collection plane hit clusters adjacent to nonfunctional wires are vetoed.

Blips are evaluated to identify candidate \BetaBi deposits, requiring a match in at least two planes.  A fiducial requirement in the $y$-$z$ plane (\mbox{$-80<y<80$~cm}, \mbox{$50<z<985$~cm}) excludes energy deposits near the edges of the active volume where space-charge distortion effects and radiological backgrounds from support struts are more prominent~\cite{uB_mev}. To reject noise and blips from $^{39}$Ar $\beta$ decays ($Q_{\beta}=0.57$~MeV), as well as high-energy blips not consistent with the $Q_{\beta}$ of $^{214}$Bi decay, we select only \BetaBi candidates with an integrated charge corresponding to energies between 0.5~MeV and 3.5~MeV.

\begin{figure}
\centering
\includegraphics[trim = 0.5cm 0cm 0cm 0cm, clip=true, width=0.95\columnwidth]{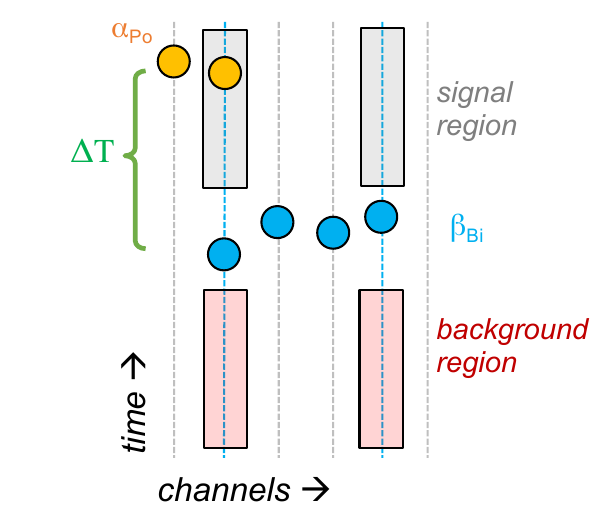}
\caption{Schematic illustrating the selection regions on the collection plane for a BiPo decay candidate. Dashed lines represent readout channels, and hits are represented as circles.  Drawing is not to scale.}
\label{fig:cartoon}
\end{figure}

After a \BetaBi candidate blip is identified, which typically spans one to four wires, we search for associated \AlphaPo candidates. As illustrated in Fig.~\ref{fig:cartoon}, the outer-most wires of the $\beta$ cluster on the collection plane are searched, one of which is assumed to correspond physically to the origin of the $\beta$'s trajectory and thus the location of the at-rest $^{214}$Po isotope.
Clusters occurring on these wires within a ``signal region'' time window of 20--500~\mus following the \Beta candidate are evaluated as potential candidates. The minimum of 20~\mus is imposed to ensure the $\alpha$ produces a distinct and well-separated signal on the readout wire. 
Only clusters with $<$~6,000 electrons are selected as candidates for the highly-quenched  \AlphaPo signal, corresponding to an electron-equivalent energy $<$~0.24~MeVee.

\subsection{Background subtraction} \label{sec:background-sub}

The time separation $\Delta$T is stored for each BiPo candidate. Such a distribution can be fit to an exponential function, with  mean decay time fixed to the known \Po lifetime, $\tau = T_{1/2}/\ln(2) = 237.0$~$\mu$s. This fit can then be used to infer the true signal content in the sample. However, our sample will be contaminated by several sources of background outlined below.

\begin{enumerate}
    \item Random electronics noise resulting in a time-independent contribution to $\Delta$T.
    
    \item Unrelated radiological or cosmic activity, such as from $\gamma$ rays or neutrons. Such topologies create groups of closely spaced blips with separations on the order of several centimeters,  leading to a $\Delta$T contribution with a characteristic time of \mbox{$\approx$~10--30~\mus}.
    
    \item Low-energy $\gamma$ rays emitted in the $\beta$ decay of radiological isotopes, including but not limited to \Bi. This background is particularly problematic since the spatial distribution of these $\gamma$ interactions relative to the \BetaBi candidate translates to a time distribution with a characteristic time constant resembling that of true BiPo decays.
\end{enumerate}

To account for these backgrounds, we repeat the selection procedure on the same wires but in a time window preceding each \BetaBi candidate, illustrated by the red-shaded box in Fig.~\ref{fig:cartoon}. Spatial symmetry with respect to the signal region ensures that the distribution of false \AlphaPo candidates, due to noise or $\gamma$ activity, will be identical to that in the forward signal region. 
Figure~\ref{fig:dt_overlay} shows the distribution of candidate decay times for the forward signal region and time-reversed background region for data taken during the Rn-doping period. 
Fitting the background region's distribution to a function modeling the three background categories discussed above suggests the approximate relative contributions of each are about 50\% \emph{(1)}, 20\% \emph{(2)}, and 30\% \emph{(3)}, respectively. 

\begin{figure}
\includegraphics[trim = 0.0cm 0cm 0.3cm 0cm, clip=true, width=0.97\columnwidth]{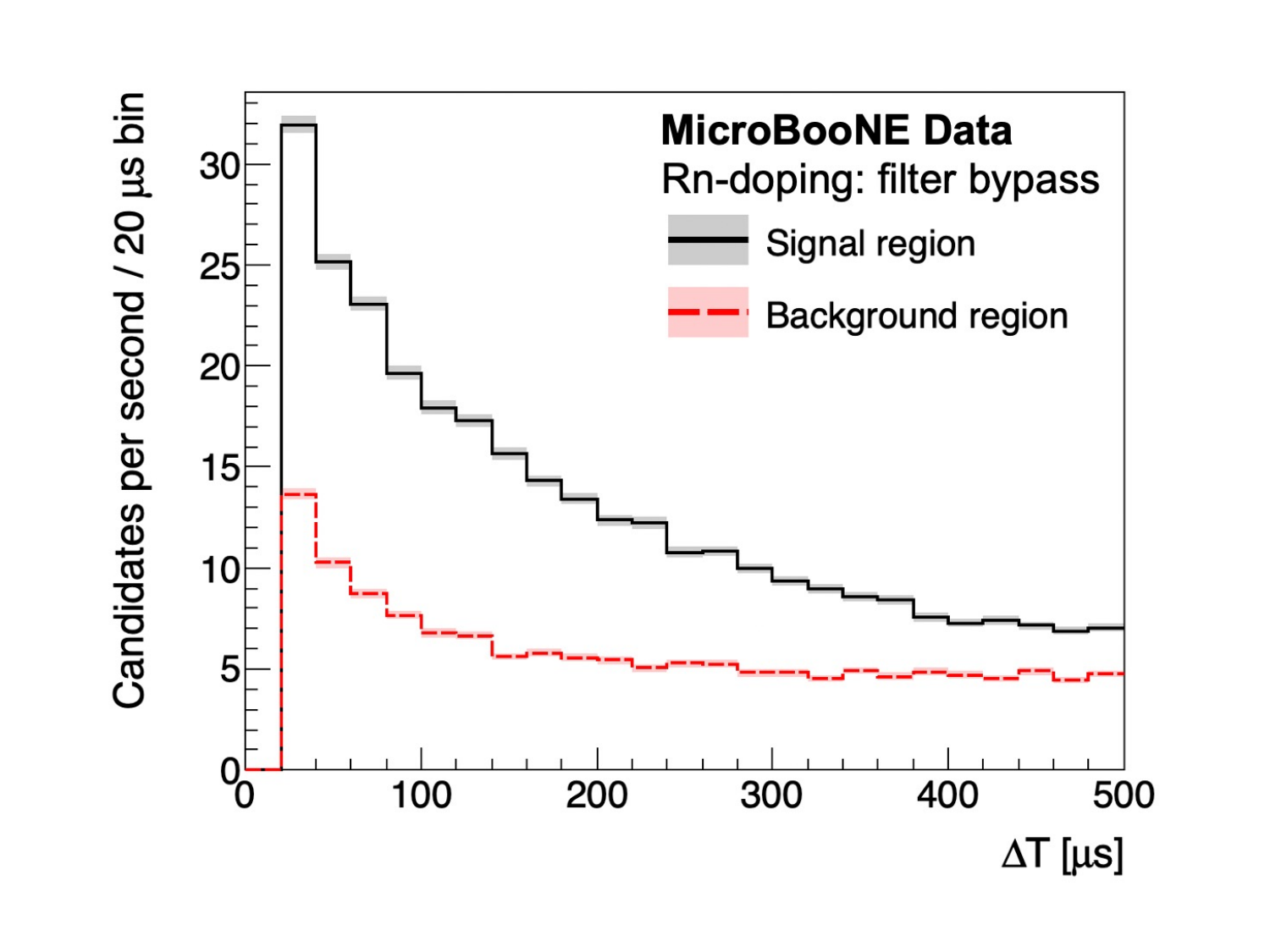}
\caption{Distributions of $\Delta$T for \BiPo candidates in the signal and background selection regions for a period during which \Rn was actively being added to the LAr.}
\label{fig:dt_overlay}
\end{figure}

We also consider additional detector effects that may influence the quantity and spatial distributions of candidates in the signal and background regions. The accumulation of slowly drifting positive Ar ions from cosmic rays distorts the electric field within the active volume, leading to a slightly higher field strength in regions nearer to the cathode and a lower field strength nearer to the anode~\cite{Abratenko_2020}. Since recombination depends on the local electric field, energy deposited nearer to the cathode (i.e., in the signal region) will produce more free charge relative to deposits nearer to the anode (i.e., in our time-reversed background region). Electron drift attenuation and diffusion will have an opposite effect, decreasing the detection efficiency for ionization in the signal region relative to the background region. We employ a data-driven method to account for the confluence of these two effects by running the selection in a ``control region'' of the collection plane separated from the \BetaBi candidate by at least several wires. Here, we expect symmetrically distributed contributions from $\gamma_\text{Bi}$ production in both the forward and backward regions, so any differences can be attributed to the aforementioned detector effects. We fit a linear function to the forward-to-backward candidate ratio per time bin and apply this as a bin-by-bin correction factor on the background region distribution. 
The end result is a downward scaling in the range of 2\%--3\% on the background distribution for normal data-taking conditions, with bins at higher $\Delta$T requiring a larger correction as expected.

\subsection{Extracting the decay rate}
\begin{figure}
\includegraphics[width=0.96\columnwidth]{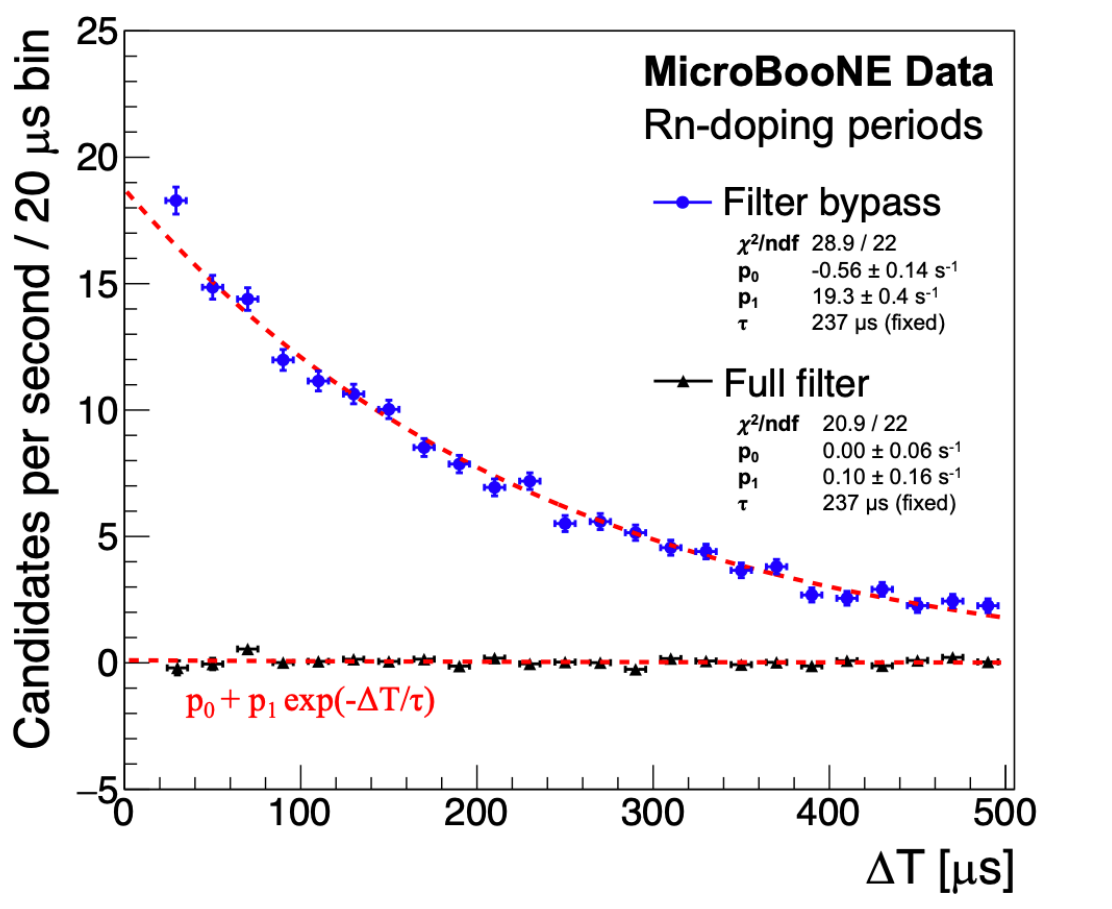}
\caption{Background-subtracted and fitted $\Delta$T distributions for the Rn-doping data for a period when the filter was bypassed (blue) and the preceding period where the full filtration system was employed (black). The lifetime  $\tau$ in both fits is fixed to the known \Po mean lifetime of 237~$\mu$s.}
\label{fig:dt_fit}
\end{figure}

The background-subtracted distribution of BiPo candidates' $\Delta$T is shown in Fig.~\ref{fig:dt_fit} for the filter bypass period and the equal-length period preceding it in which the full filter was employed. Both subtracted distributions are well-described by a fitted function of the form
$p_0 + p_1\exp{(-{\Delta}T/\tau)}$, 
where $\tau$ is fixed\footnote{
When all three parameters are allowed to vary freely, the filter bypass data yields a best-fit lifetime of $\tau \approx 200 \pm 15$~$\mu$s, roughly consistent with the known \Po lifetime. 
} to the known \Po lifetime of 237~$\mu$s~\cite{TOTH1977437}.
The fit to the filter bypass data exhibits a prominent exponential component compared to the full-filter data, indicating the presence of a population of BiPo decays among the selected candidates. 
Integrating the exponential component of the fit function allows us to extrapolate the rate of reconstructed BiPo decays present in the data regardless of the length of the time window used to select candidates. 
In the nominal fit, the constant background term $p_0$ is treated as a free parameter to account for the possibility of a background subtraction imperfection.  To account for this uncertainty, we repeat the fit with $p_0$ fixed to zero and treat the difference in outcome between this and our nominal fit as a systematic uncertainty.
This procedure yields an average rate of $0.73\pm0.05$~BiPo decay candidates per 3.2~ms TPC readout period within the reduced fiducialized volume for the filter bypass data,  compared to $(4\pm6)\times10^{-3}$ per readout when the full filter was in use.

\begin{figure}
\includegraphics[trim = 0.2cm 0cm 0.3cm 0cm, clip=true, width=0.98\columnwidth]{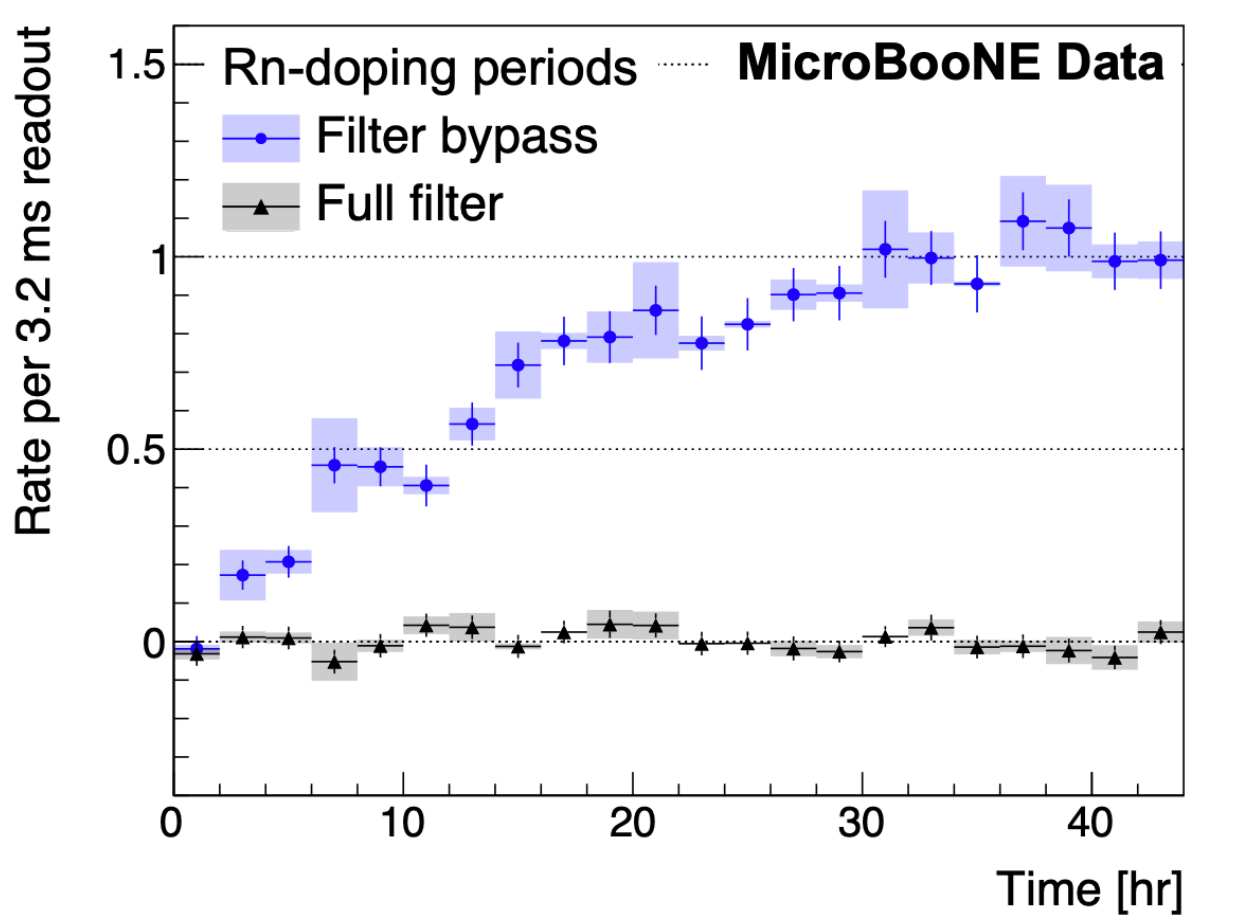}
\caption{Measured rate of BiPo decays per readout within the fiducial volume, plotted against the time of the event relative to the start of each Rn-doping data-taking period. Statistical errors are represented by solid lines, while systematic errors from uncertainties in the fit methodology as described in the text are represented by shaded regions.}
\label{fig:rate_vs_time}
\end{figure}
To visualize the time evolution of these measurements, we divide the data into 2-hour periods and perform this technique in each of them separately. The resulting rates as a function of are shown in Fig.~\ref{fig:rate_vs_time}. Vertical error bars include contributions from both the returned fit uncertainty and the systematic uncertainty from fixing the fit parameter $p_0$. 

}

\section{Monte Carlo simulation} \label{sec:simulation}
{

\subsection{Generated samples}

To translate a measured BiPo rate per TPC readout window into a measurement of the activity of $^{222}$Rn in MicroBooNE's liquid argon, the efficiency of the BiPo selection described in Sec.~\ref{sec:analysis} must be  corrected for.  Monte Carlo (MC) simulations are used to characterize this efficiency. With the aid of the \texttt{Decay0} radioactive decay generator~\cite{Ponkratenko:2000um}, a list of $\gamma$ and $\beta$ rays are generated matching the kinematic and time distributions expected from individual correlated \BiPo decay pairs. These particle lists are used to generate simulated MicroBooNE events, each containing 40~decays distributed randomly throughout the active volume within a time of $\pm$2.8~ms relative to the start of the main drift window. 
Particle propagation and detector readout are simulated using an integration of the LArSoft~\cite{larsoft} the Geant4~\cite{g4} software packages referred to as ``LArG4.'' To realistically account for cosmic backgrounds and for electronics noise present in data, which are challenging to accurately model in simulation, wire signals from each simulated event are overlaid onto an unbiased beam-external data event. Each overlaid event is then processed by the reconstruction and signal selection.

\begin{table}[tb]
    \centering
    \begin{tabular}{ l  c }
    \hline \hline
    {Simulation parameter} & {Setting}\\
    \hline
    Average electric field & 274~V/cm \\
    Recombination model ($e^{\pm}$) & Modified box~\cite{argo_modbox} \\
    Electron drift speed & 1.1~mm/$\mu$s~\cite{ub_efield} \\
    Longitudinal diffusion, $D_L$   & 3.74~cm$^2$/s~\cite{ub_diffusion}\\
    Transverse diffusion, $D_T$ & 5.85~cm$^2$/s~\cite{ub_diffusion} \\
    \hline \hline
    \end{tabular}
    \caption{Selected parameters used in the MC simulations that are most impactful on MeV-scale reconstruction capabilities.}
    \label{tab:sim_parameters}
\end{table}

Table~\ref{tab:sim_parameters} summarizes the crucial detector physics parameters used to generate this MC dataset. In the LArG4 framework, electron-ion recombination is simulated with the modified box model mentioned in Sec.~\ref{sec:reconstruction_energy}.
Since the ArgoNeuT Collaboration used data from stopping protons and deuterons to parameterize this model, it is applicable for $dE/dx < 35$~MeV/cm~\cite{argo_modbox}. 
For $\alpha$ particles and nuclear recoils, which are more highly-ionizing, additional charge quenching effects must be considered~\cite{MEI200812,HITACHI2005247,Hitachi:2004zw}. 
In this analysis, the charge deposited by $\alpha$ particles comes from an empirical field-dependent 
model based on fits to existing data, crafted by the Noble Element Simulation Technique (NEST) Collaboration ~\cite{nest_paper,nest_software}. A random Poisson-like smearing is then applied to the ionization yield ($\sigma = \sqrt{N_e}$) to mimic binomial fluctuations.
This approach predicts a mean $\alpha$ charge-yield (QY) of  about 390~$e^-$/MeV compared to the modified box model's prediction of 530~$e^-$/MeV. 

Sources of physics-related systematic uncertainty are studied using additional samples with key simulation parameters varied accordingly. The dominant source of systematic uncertainty is the $\alpha$ QY. Since the few existing $\alpha$ data in LAr do not report measurement errors, NEST assigns a $\pm$10\% uncertainty on its empirical model. We assume an uncertainty of $\pm$20\% for this analysis. 

Electron drift diffusion is particularly impactful for low-energy deposits in LAr. Since these features typically span only a few wires, any charge within the main electron cloud that diffuses far enough to be collected on neighboring wires is less likely to produce signals above threshold. The value for the longitudinal diffusion simulated in this analysis comes from a recent MicroBooNE measurement of $D_L~=~3.74^{+0.28}_{-0.29}$~cm$^2$/s~\cite{ub_diffusion}. This analysis also predicts the associated transverse diffusion, $D_T$, though no direct measurement of $D_T$ exists at MicroBooNE's electric field. Systematic samples are generated with correlated variations in $D_L$ and $D_T$ of $\pm1\sigma$ and $\pm$30\%, respectively.

Systematic effects from MicroBooNE's calibrated energy scale ($e^-$ per ADC) are addressed through samples in which all charge deposits are scaled up or down by 5\%. Recombination modeling uncertainties are addressed by using an alternative parametrized model~\cite{lar_diff2} and by enhancing recombination fluctuations by a factor of 10 as some data suggest~\cite{APRILE1987519}.

}

\subsection{Calorimetric validation}
{

\begin{figure}[tb]
\includegraphics[trim = 0.0cm 0cm 0.0cm 0cm, clip=true,width=0.98\columnwidth]{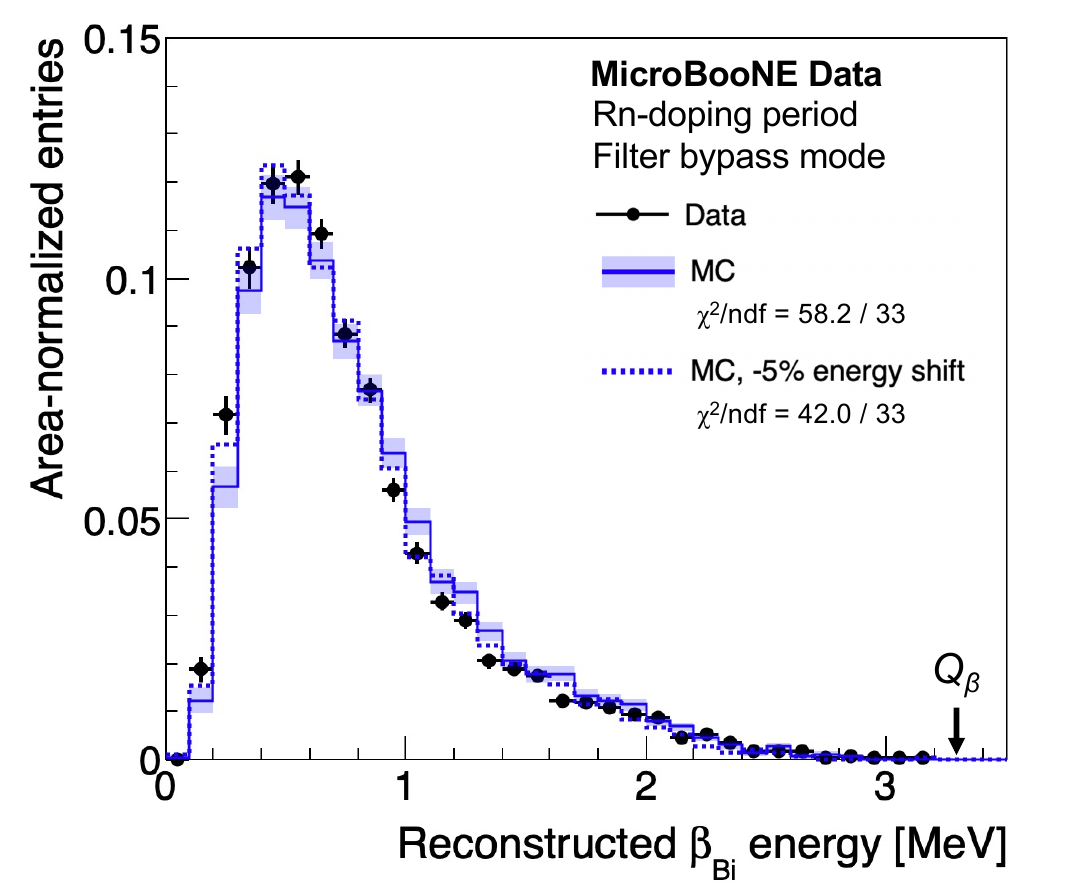}
\caption{The reconstructed \BetaBi energy spectrum. The shaded region represents MC statistical uncertainty, while the blue dotted line is the MC spectrum with a -5\% energy scale shift. The value of $Q_{\beta} = 3.27$~MeV for \Bi is indicated.}
\label{fig:beta}
\end{figure}

\begin{figure}
\includegraphics[trim = 0.0cm 0cm 0.0cm 0cm, clip=true, width=0.97\columnwidth]{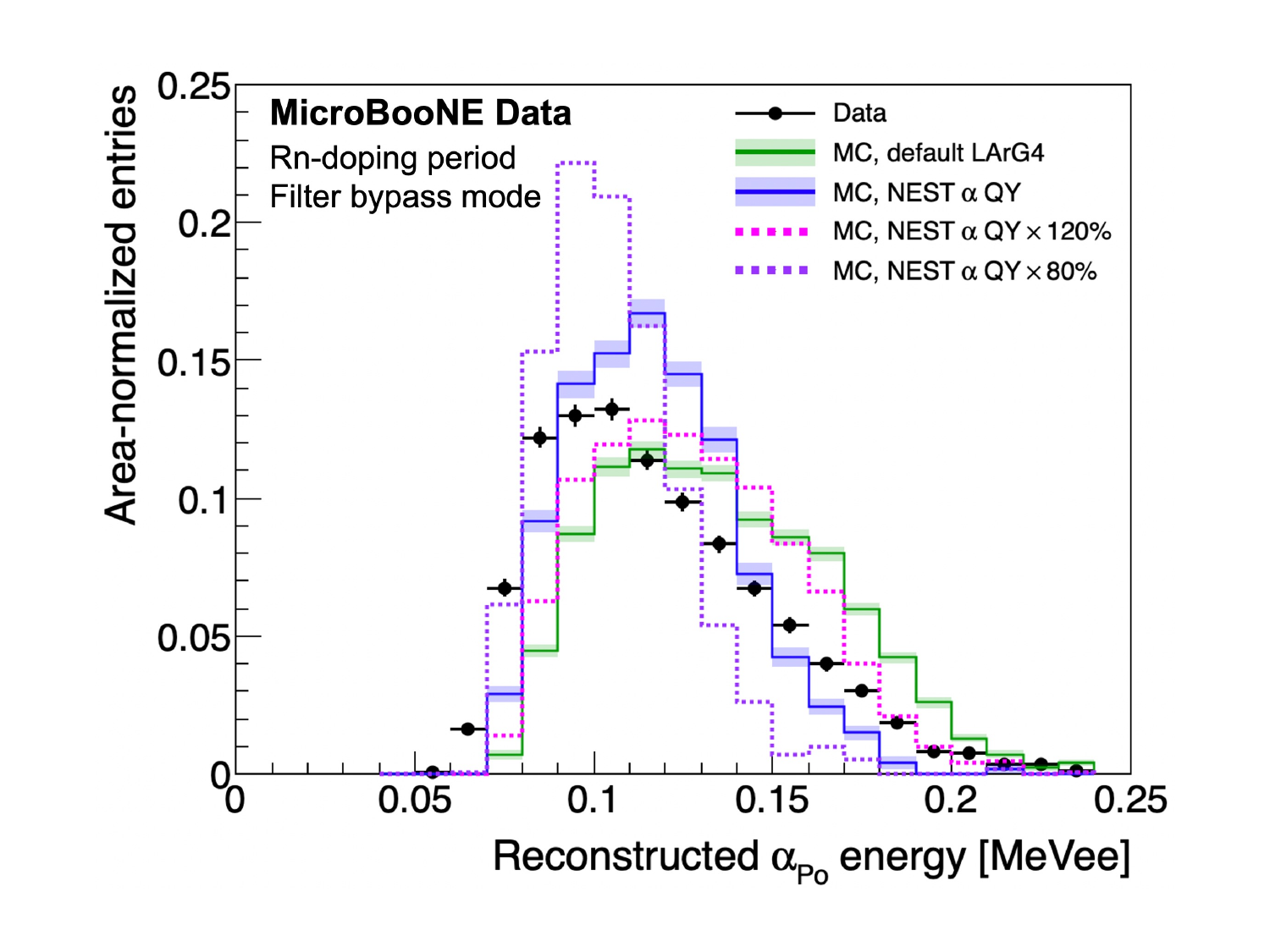}
\caption{Reconstructed \AlphaPo energy spectrum, in electron-equivalent units, following the background-subtraction procedure. Due to the uncertainty in the $\alpha$ QY in LAr, additional samples are generated using NEST's empirical model~\cite{nest_software} with a $\pm$20\% scaling applied to the QY. The LArG4 MC, which by default uses particle $dE/dx$ from Geant4~\cite{g4} as input to the modified box model to calculate recombination, is shown for comparison.}
\label{fig:alpha}
\end{figure}
While precise calorimetry is not essential for \BiPo signal selection, energy spectra are reconstructed to validate the simulation of low-energy signatures.  
These validations further extend the demonstrated boundaries of charge-based reconstruction capabilities in large single-phase LArTPCs.
Energy reconstruction follows the procedure laid out in Sec.~\ref{sec:reconstruction_energy}, allowing us to translate collected charge into ``electron-equivalent'' energy using Eq.~\ref{eq:electronEquivEnergy} in which an electronlike recombination factor is assumed.
A similar background subtraction technique as described in Sec.~\ref{sec:analysis} is performed on the energy distributions of \BetaBi and \AlphaPo candidates using information from the collection plane.
The filter bypass Rn-doping dataset is used for these calorimetric checks. Data was excluded beyond 35~hours when the measured drift electron lifetime was found to drop below $\approx7$~ms as LAr purity decreased. A corresponding MC sample was generated with an electron lifetime of 8~ms to match the average level of attenuation observed in data events with tagged BiPo candidates.

The same background subtraction procedure described  in Sec.~\ref{sec:background-sub} is now applied to the distribution of reconstructed \BetaBi energies. Figure~\ref{fig:beta} shows this background-subtracted spectrum for data and MC simulation, with the usual energy-based selection requirement \mbox{($E_\beta > 0.5$~MeV)} dropped to reveal the full spectrum. As expected, the data exhibit a tail extending out to \mbox{$\approx3.3$~MeV} matching the $Q_\beta$ value of \Bi. The shape of the lower end of the spectrum is sculpted by energy threshold effects discussed in Sec.~\ref{sec:reconstruction_geo}; the efficiency for reconstructing plane-matched blips drops rapidly for electron energies below 0.7~MeV, reaching 50\% around 0.5~MeV and becoming negligible by \mbox{$\approx0.1$~MeV.} A goodness-of-fit test between data and the MC yields a $\chi^2$ of 58 over 33 degrees of freedom (ndf). Applying an energy shift of $-5\%$ to the MC  (equivalent to the calibrated energy scale uncertainty) improves the match, yielding $\chi^2$/ndf  = 42/33.

The reconstructed \AlphaPo energy spectrum from the filter-bypassed Rn-doping R\&D run period is shown in Fig.~\ref{fig:alpha}. 
Since the 7.7~MeV $\alpha$ particle experiences significant charge quenching in LAr, its reconstructed energy in electron-equivalent units ranges from only 50~keV to 200~keV.  
Unlike for the \BetaBi signal, the selection of the correlated \AlphaPo signal takes place entirely on the collection plane with no plane-matching requirements imposed. 
As shown in Fig.~\ref{fig:reco_efficiency} and Table~\ref{tab:reco_efficiency}, the reconstruction efficiency extends far lower in energy on the collection plane alone compared to when plane-matching requirements are imposed. 
Despite this lowered threshold, the reconstructed \AlphaPo spectrum occupies the very lowest extent of the sensitivity, with an average hit-finding efficiency of $\approx10$\% in the 100--150~keV true electron-equivalent energy range encompassing the \AlphaPo signal, and \mbox{$<2\%$} below 100~keV true energy.  
The shape of the spectrum is heavily sculpted by this sudden turn-on in sensitivity, exhibiting a sharp rising edge from 70--90~keVee.  
This same thresholding effect is also visible in the MC samples, though offset from data by slightly less than 10~keV.
With the $\alpha$ QY scaled up by 20\%, the distribution skews too high, overshooting the high-energy tail of the data and resulting in a softened rising edge at the lower end.
When the $\alpha$ QY is scaled down by 20\%, the high-energy tail does not extend out as far as the data and the rising edge sharpens. 
While there is some broad qualitative agreement in the \AlphaPo spectrum between data and MC, this comparison highlights the unresolved systematic uncertainties in modeling this signal.

\subsection{Efficiency} 
\label{sec:reconstruction_eff}

Since we demonstrated the accuracy of the simulation through data-MC calorimetric comparisons, we now use it to determine the efficiency in measuring the rate of \BiPo decays. To best reflect standard MicroBooNE operating conditions, the simulated drift electron lifetime is set sufficiently high such that charge attenuation is negligible.
The analysis procedure is carried out on each MC sample and the underlying cosmic data overlaid onto the simulated events. 
The overlay data by itself yields a measured rate of $0.02 \pm 0.02$~candidates per readout. This is subtracted off the rates obtained from each MC overlay sample in order to properly isolate the efficiency of the MC contribution in each.

For the nominal MC sample, a rate of $1.18\pm0.13$ decays per readout is measured compared to the simulated rate within the limited fiducial region of 14.2~decays per readout, resulting in an efficiency of \mbox{$\epsilon_\text{nom} = (8.3\pm0.9)$\%}. Effects due to nonfunctional wires, thresholding, vetoing of hits surrounding cosmic tracks, blip candidate selection cuts, and the background subtraction procedure 
are folded into this efficiency. The uncertainty on $\epsilon_\text{nom}$ arises primarily from the systematic uncertainty assigned during the fitting procedure described in Sec.~\ref{sec:analysis}. 
Table~\ref{tab:systematics} reports the relative impact on MC efficiency for each physics-related source of systematic uncertainty. Uncertainties related to the $\alpha$ QY and electron diffusion dominate the error budget. Added in quadrature, the total systematic uncertainty on efficiency is about $\pm$50\%, yielding a final efficiency of $\epsilon = (8.3 \pm 4.2)$~\%.

\begin{table}[tb]
    \centering
     \begin{tabular}{ l  c  }
     \hline \hline
      Systematic
        & Uncertainty \\
        \hline   
        Alpha QY
        & $\pm$43\% \\ 
        Electron diffusion
        & +26\%, -17\% \\  
        Energy scale
        & $\pm$15\% \\
        Recombination modeling
        & $\pm$1.9\% \\
        \hline 
        \makecell[l]{Total}
        & +52\%, -49\% \\
        \hline \hline
    \end{tabular}
    \caption{Summary of physics-related systematic effects considered in this analysis, along with their relative impact on the MC-derived efficiency ($\delta\epsilon/\epsilon_{\text{nom}}$). The bottom row includes the total quadrature sum of all effects listed. 
    }
    \label{tab:systematics}
    \vspace{2ex}
\end{table}

}

\section{Results and discussion} 
\label{sec:results}
{
To measure the ambient rate of \Bi decays in standard MicroBooNE operating conditions, rather than during  R\&D periods used in previous sections during which radon was actively being added to the TPC, we use a sample of unbiased beam-external events acquired over a period of nearly seven weeks during the 2018 physics data-taking campaign described in Sec.~\ref{sec:microboone}. Figure~\ref{fig:run3_dt_fit} shows the background-subtracted $\Delta$T distribution for these data, fitted to the exponential function $f = p_0 + p_1\exp{(-{\Delta}T/\tau)}$, with $\tau$ fixed to the \Po lifetime. 
Integrating the BiPo component of the fit and incorporating statistical and systematic uncertainties from Sec.~\ref{sec:analysis}, a rate of $(0.2 \pm 2.3) \times 10^{-3}$ 
candidates per readout is obtained within the inner fiducial volume defined by our $yz$-plane selection cuts described in Sec.~\ref{sec:analysis_sigselection}.
The error on this rate is dominated by the statistical uncertainty from the fit.

\begin{figure}
\begin{center}
\includegraphics[trim = 0.3cm 0cm 0.0cm 0cm, clip=true,width=0.98\columnwidth]{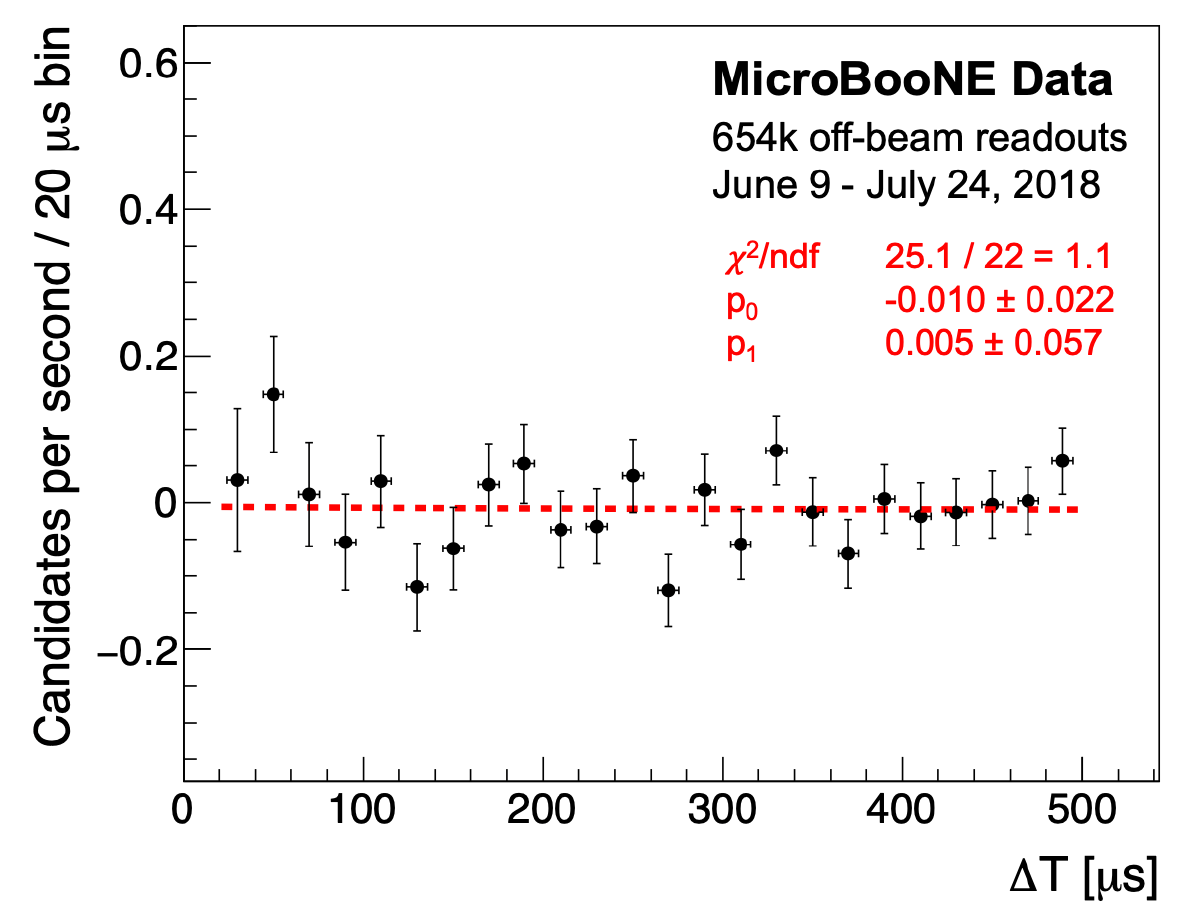}
\caption{The fitted $\Delta$T distribution from unbiased nonbeam data taken during a 46-day period of standard operating conditions in MicroBooNE.}.
\end{center}
\label{fig:run3_dt_fit}
\end{figure}

This rate is converted to a measurement of activity by correcting for the MC efficiency ($\epsilon$) found in Sec.~\ref{sec:reconstruction_eff} and dividing by the total mass of LAr in the limited fiducial volume that was sampled. This yields a measured radioactive \Bi activity of $(0.01 \pm 0.16 \text{(stat)} \pm 0.06 \text{(syst)})$~mBq/kg =~\mbox{$(0.01 \pm 0.17)$ mBq/kg}. Given that this result is consistent with zero, an upper limit of \mbox{$<0.35$~mBq/kg} is placed at the 95\% confidence level.

\begin{figure}
\includegraphics[trim = 0cm 0cm 0cm 0cm, clip=true, width=0.98\columnwidth]{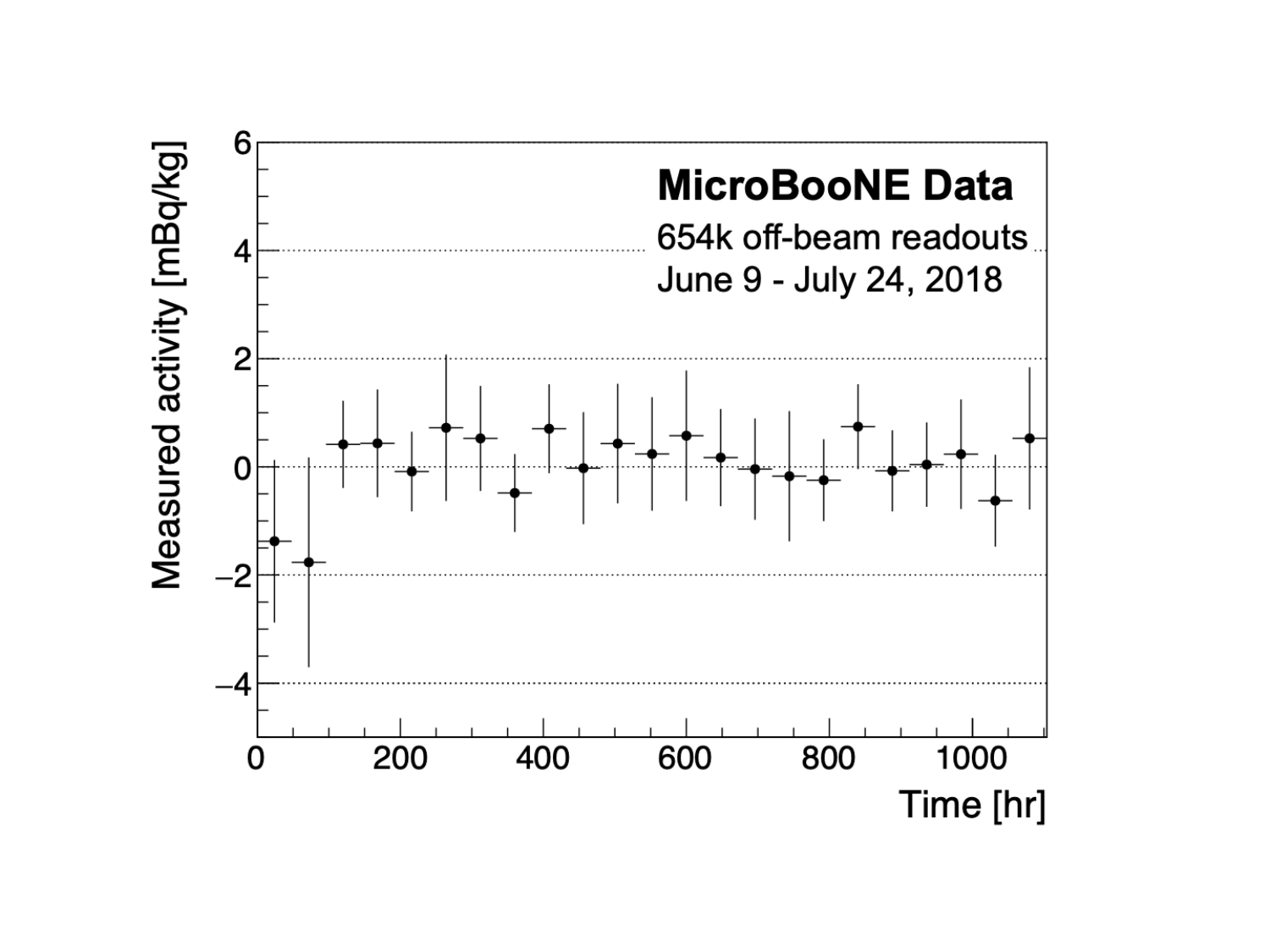}
\caption{The efficiency-corrected BiPo rate measured in 48-hour periods throughout the beam-external dataset. Error bars on each data point are dominated by statistical uncertainties.}
\label{fig:run3_rate_vs_time}
\end{figure}

The dataset is divided into a series of 48-hour periods, and the \Bi rate measurement is repeated in each.  A lower unbiased trigger rate is used in normal data-taking compared to the R\&D runs used previously, necessitating a longer time period to achieve sufficient per-bin statistics.  Rates for each period are shown in Fig.~\ref{fig:run3_rate_vs_time}.  No major trends are observed that would indicate sudden changes in the LAr circulation system's operational state or gradual degradation in filter efficiency.

To relate the measured activity of \Bi to that of ambient \Rn, we must consider the impact of so-called ``plate-out'' effects observed in LXe detectors~\cite{XENON_RnKr, PhysRevD.108.012010, next_radon} and in LAr detectors like DEAP-3600~\cite{PhysRevD.100.022004}. This effect arises as isotopes produced in a positive charge state are drifted toward the cathode or brought into contact with the field cage walls through convective fluid motion, where they then attach to the material, thus reducing the measurable activity of radon progeny lower in the decay chain.

Since plate-out is not simulated in MicroBooNE, we estimate the magnitude of this effect with a toy model simulation. We assume an initial homogeneous distribution of \Rn 
and neglect the possible neutralization of ions by drifting electrons from cosmic ray ionization activity. While the velocity of the LAr convective flow is similar to the ion drift speed of several mm per second~\cite{Abratenko_2020}, we assume no net preferred flow direction over long timescales, and therefore neglect this effect in our model. For each \Rn, the cascade of subsequent decays is simulated,  
with each daughter randomly assigned a positive or neutral charge state based on isotope-specific measurements.  The measured ion fraction of ($37\pm3$)\% is used for $^{218}$Po~\cite{darkside50_ionfrac}. Corresponding measurements in LAr for other progeny do not exist, so we assume the same ion fraction of 37\% for $^{214}$Pb, and estimate 56\% for \Bi by assuming that the ratio of the measured ion fractions for $^{218}$Po between LAr and LXe apply as well for the \Bi isotope, which has been measured only in LXe~\cite{exo200_ionfrac}. Produced ions are drifted a random distance toward the cathode based on the isotope's known decay lifetime, using the drift speed measured in LAr for $^{218}$Po$^+$
equivalent to 0.23~cm/sec at MicroBooNE's electric field strength~\cite{darkside50_ionfrac}.
If an ion reaches the cathode, it and all its progeny remain permanently plated.   
Due to the challenge of simulating isotopes attached or embedded onto surfaces, it is not known whether their $\alpha$ and $\beta$ decay products still produce observable signals in the LAr. 
For this rough estimate, we consider both limiting cases (50\% observable and 0\% observable), resulting in a ratio between the \Rn activity and the measurable \BiPo rate of $R_\text{Rn}/R_\text{BiPo} \approx 2.3 \pm 0.4$ for the MicroBooNE active volume.

Using this ratio, the measured \Bi activity corresponds to an estimated ambient \Rn activity of 
\mbox{$\approx (0.03 \pm 0.39)$~mBq/kg} 
in MicroBooNE's bulk LAr. 
This level of contamination likely satisfies the \Rn radiopurity target for DUNE's  low-energy physics program of $<1~\text{mBq/kg}$~\cite{Avasthi:2022tjr}. 
Given the similarity in LAr filtration system design and components between MicroBooNE and DUNE~\cite{DUNE:2020txw, dune_cdr_2023}, we should expect similar radon levels in DUNE's bulk LAr if a comparable cryogenic recirculation period can be achieved.  MicroBooNE operated with a LAr volume exchange period of about 2.5~days~\cite{ub_det}. DUNE, with its vastly larger LAr volumes, currently expects an initial volume exchange period of 5.5~days, gradually slowing to 11~days for long-term operations~\cite{dune_cdr_2023}. 

The result from this analysis lacks the precision necessary for direct relevance to next-generation dark matter experiment radiopurity goals. However, when combined with Ref.~\cite{ub_radon}, this result suggests promising intrinsic capabilities of liquid-phase filtration systems for achieving high radiopurities, which should be further investigated in liquid noble element dark matter R\&D efforts.  Analyses with higher statistical precision and lower inherent background contamination should be performed with future Fermilab-based LArTPCs such as SBND~\cite{sbnd_phys}, given its larger LAr volume and highly capable light collection system. 
}

\section{Conclusion} \label{sec:conclusion}
{
Using the MicroBooNE charge collection system and newly developed low-energy reconstruction tools, we have probed the presence of \Rn  in a large LArTPC by identifying MeV-scale energy depositions produced in decays of its progeny isotopes~\Bi and~\Po.  
Blips matching the expected appearance of~\Bi decay $\beta$ particles were identified and reconstructed using a multiplane scheme. Weaker blips matching the appearance of subsequent \Po decay $\alpha$ particles were then reconstructed in a narrow region of spatial/temporal phase space with respect to the~\Bi signal.  Backgrounds to coincident \BiPo signals arising from randomly-coincident blips, multisite $\gamma$ ray interactions, and $\beta+\gamma$ radon progeny decays were subtracted using off-window and time-reversed-window sideband methods. 
By estimating the efficiency for signal detection using MC simulations and validating these simulations with special MicroBooNE R\&D datasets, measured \BiPo rates were reliably converted into measurements of radioactive bismuth activity. 

We do not detect any presence of \Bi in steady-state MicroBooNE physics data-taking conditions, and place a limit of \mbox{$<0.35~\text{mBq/kg}$} at the 95\% confidence level with measurement errors dominated by statistical uncertainties.
Based on a toy simulation that extrapolates the rate of \Bi to that of \Rn in the LAr bulk, we estimate a corresponding radon activity that
satisfies the targeted upper limit for the DUNE LArTPC experiment's baseline low-energy physics program of \mbox{$<1$~mBq/kg}. This was achieved by MicroBooNE in the absence of any direct efforts towards radio-purification.  This also represents the first \emph{in situ} measurement of bulk radiopurity in a LAr particle detector employing liquid filtration. 

In performing this measurement, we have extended the  boundaries of charge-based calorimetry and reconstruction capabilities in large single-phase neutrino LArTPCs. We accurately reconstruct the energy spectrum of $\beta$ particles in \Bi decay within an energy range of 0.2--3.0~MeV, and identify and reconstruct \Po decay $\alpha$ particles with 75--200~keV of electron-equivalent energy.  
To our knowledge, these are the lowest energies at which particle calorimetry and identification capabilities have been demonstrated so far in a single-phase neutrino LArTPC.  
}

\begin{acknowledgments}
{
This document was prepared by the MicroBooNE Collaboration using the resources of the Fermi National Accelerator Laboratory (Fermilab), a U.S. Department of Energy, Office of Science, HEP User Facility. Fermilab is managed by Fermi Research Alliance, LLC (FRA), acting under Contract No. DE-AC02-07CH11359.  MicroBooNE is supported by the following: the U.S. Department of Energy, Office of Science, Offices of High Energy Physics and Nuclear Physics; the U.S. National Science Foundation; the Swiss National Science Foundation; the Science and Technology Facilities Council (STFC), part of the United Kingdom Research  and Innovation; the Royal Society (United Kingdom); and the UK Research  and Innovation (UKRI) Future Leaders Fellowship. Additional support for the laser calibration system and cosmic ray tagger was provided by the Albert Einstein Center for Fundamental Physics, Bern, Switzerland. We also acknowledge the contributions of technical and scientific staff to the design, construction, and operation of the MicroBooNE detector as well as the contributions of past collaborators to the development of MicroBooNE analyses, without whom this work would not have been possible. For the purpose of open access, the authors have applied a Creative Commons Attribution (CC BY) public copyright license to any Author Accepted Manuscript version arising from this submission.
}
\end{acknowledgments}

\bibliography{refs}

\begin{thebibliography}{84}%
\makeatletter
\providecommand \@ifxundefined [1]{%
 \@ifx{#1\undefined}
}%
\providecommand \@ifnum [1]{%
 \ifnum #1\expandafter \@firstoftwo
 \else \expandafter \@secondoftwo
 \fi
}%
\providecommand \@ifx [1]{%
 \ifx #1\expandafter \@firstoftwo
 \else \expandafter \@secondoftwo
 \fi
}%
\providecommand \natexlab [1]{#1}%
\providecommand \enquote  [1]{``#1''}%
\providecommand \bibnamefont  [1]{#1}%
\providecommand \bibfnamefont [1]{#1}%
\providecommand \citenamefont [1]{#1}%
\providecommand \href@noop [0]{\@secondoftwo}%
\providecommand \href [0]{\begingroup \@sanitize@url \@href}%
\providecommand \@href[1]{\@@startlink{#1}\@@href}%
\providecommand \@@href[1]{\endgroup#1\@@endlink}%
\providecommand \@sanitize@url [0]{\catcode `\\12\catcode `\$12\catcode
  `\&12\catcode `\#12\catcode `\^12\catcode `\_12\catcode `\%12\relax}%
\providecommand \@@startlink[1]{}%
\providecommand \@@endlink[0]{}%
\providecommand \url  [0]{\begingroup\@sanitize@url \@url }%
\providecommand \@url [1]{\endgroup\@href {#1}{\urlprefix }}%
\providecommand \urlprefix  [0]{URL }%
\providecommand \Eprint [0]{\href }%
\providecommand \doibase [0]{https://doi.org/}%
\providecommand \selectlanguage [0]{\@gobble}%
\providecommand \bibinfo  [0]{\@secondoftwo}%
\providecommand \bibfield  [0]{\@secondoftwo}%
\providecommand \translation [1]{[#1]}%
\providecommand \BibitemOpen [0]{}%
\providecommand \bibitemStop [0]{}%
\providecommand \bibitemNoStop [0]{.\EOS\space}%
\providecommand \EOS [0]{\spacefactor3000\relax}%
\providecommand \BibitemShut  [1]{\csname bibitem#1\endcsname}%
\let\auto@bib@innerbib\@empty
\bibitem [{\citenamefont {Rubbia}(1977)}]{Rubbia:1977zz}%
  \BibitemOpen
  \bibfield  {author} {\bibinfo {author} {\bibfnamefont {C.}~\bibnamefont
  {Rubbia}},\ }\href@noop {} {\bibinfo {title} {{The liquid-argon time
  projection chamber: A new concept for neutrino detectors}}} (\bibinfo {year}
  {1977}),\ \bibinfo {note}
  {\href{http://cds.cern.ch/record/117852}{CERN-EP-INT-77-8}}\BibitemShut
  {NoStop}%
\bibitem [{\citenamefont {Anderson}\ \emph {et~al.}(2012)\citenamefont
  {Anderson} \emph {et~al.}}]{Anderson:2012vc}%
  \BibitemOpen
  \bibfield  {author} {\bibinfo {author} {\bibfnamefont {C.}~\bibnamefont
  {Anderson}} \emph {et~al.},\ }\bibfield  {title} {\bibinfo {title} {{The
  ArgoNeuT detector in the NuMI low-energy beam line at Fermilab}},\ }\href
  {https://doi.org/10.1088/1748-0221/7/10/P10019} {\bibfield  {journal}
  {\bibinfo  {journal} {{J. Instrum.}}\ }\textbf {\bibinfo {volume} {7}},\
  \bibinfo {pages} {P10019} (\bibinfo {year} {2012})}\BibitemShut {NoStop}%
\bibitem [{\citenamefont {Acciarri}\ \emph
  {et~al.}(2017{\natexlab{a}})\citenamefont {Acciarri} \emph
  {et~al.}}]{ub_det}%
  \BibitemOpen
  \bibfield  {author} {\bibinfo {author} {\bibfnamefont {R.}~\bibnamefont
  {Acciarri}} \emph {et~al.} (\bibinfo {collaboration} {MicroBooNE}),\
  }\bibfield  {title} {\bibinfo {title} {{Design and construction of the
  MicroBooNE detector}},\ }\href
  {https://doi.org/10.1088/1748-0221/12/02/P02017} {\bibfield  {journal}
  {\bibinfo  {journal} {{J. Instrum.}}\ }\textbf {\bibinfo {volume} {12}},\
  \bibinfo {pages} {P02017} (\bibinfo {year} {2017}{\natexlab{a}})}\BibitemShut
  {NoStop}%
\bibitem [{\citenamefont {Acciarri}\ \emph {et~al.}(2019)\citenamefont
  {Acciarri} \emph {et~al.}}]{argo_mev}%
  \BibitemOpen
  \bibfield  {author} {\bibinfo {author} {\bibfnamefont {R.}~\bibnamefont
  {Acciarri}} \emph {et~al.} (\bibinfo {collaboration} {ArgoNeuT}),\ }\bibfield
   {title} {\bibinfo {title} {{Demonstration of MeV-scale physics in liquid
  argon time projection chambers using ArgoNeuT}},\ }\href
  {https://doi.org/10.1103/PhysRevD.99.012002} {\bibfield  {journal} {\bibinfo
  {journal} {Phys. Rev. D}\ }\textbf {\bibinfo {volume} {99}},\ \bibinfo
  {pages} {012002} (\bibinfo {year} {2019})}\BibitemShut {NoStop}%
\bibitem [{\citenamefont {Acciarri}\ \emph
  {et~al.}(2020{\natexlab{a}})\citenamefont {Acciarri} \emph
  {et~al.}}]{argo_mcp}%
  \BibitemOpen
  \bibfield  {author} {\bibinfo {author} {\bibfnamefont {R.}~\bibnamefont
  {Acciarri}} \emph {et~al.} (\bibinfo {collaboration} {ArgoNeuT}),\ }\bibfield
   {title} {\bibinfo {title} {{Improved Limits on Millicharged Particles Using
  the ArgoNeuT Experiment at Fermilab}},\ }\href
  {https://doi.org/10.1103/PhysRevLett.124.131801} {\bibfield  {journal}
  {\bibinfo  {journal} {Phys. Rev. Lett.}\ }\textbf {\bibinfo {volume} {124}},\
  \bibinfo {pages} {131801} (\bibinfo {year} {2020}{\natexlab{a}})}\BibitemShut
  {NoStop}%
\bibitem [{\citenamefont {{MicroBooNE Collaboration}}(2018)}]{uB_ar39}%
  \BibitemOpen
  \bibfield  {author} {\bibinfo {author} {\bibnamefont {{MicroBooNE
  Collaboration}}},\ }\bibfield  {title} {\bibinfo {title} {{Study of
  Reconstructed $^{39}$Ar Beta Decays at the MicroBooNE Detector}},\ }\href
  {https://www.osti.gov/biblio/1573057} {\bibfield  {journal} {\bibinfo
  {journal} {{MICROBOONE-NOTE-1050-PUB}}\ } (\bibinfo {year}
  {{2018}})}\BibitemShut {NoStop}%
\bibitem [{\citenamefont {Bhat}(2021)}]{uB_mev}%
  \BibitemOpen
  \bibfield  {author} {\bibinfo {author} {\bibfnamefont {A.}~\bibnamefont
  {Bhat}},\ }\emph {\bibinfo {title} {{MeV Scale Physics in MicroBooNE}}},\
  \href@noop {} {Ph.D. thesis},\ \bibinfo  {school} {Syracuse University}
  (\bibinfo {year} {2021}),\ \bibinfo {note}
  {{\href{https://www.osti.gov/biblio/1824656}{FERMILAB-THESIS-2021-14}}}\BibitemShut
  {NoStop}%
\bibitem [{\citenamefont {Arneodo}\ \emph {et~al.}(2000)\citenamefont {Arneodo}
  \emph {et~al.}}]{ICARUS:2000ipe}%
  \BibitemOpen
  \bibfield  {author} {\bibinfo {author} {\bibfnamefont {F.}~\bibnamefont
  {Arneodo}} \emph {et~al.} (\bibinfo {collaboration} {ICARUS}),\ }\bibfield
  {title} {\bibinfo {title} {{{The ICARUS liquid argon time projection
  chamber}}},\ }\href
  {https://doi.org/https://doi.org/10.1016/S0168-9002(01)01003-8} {\bibfield
  {journal} {\bibinfo  {journal} {Nucl. Instrum. Meth.}\ }\textbf {\bibinfo
  {volume} {A471}},\ \bibinfo {pages} {272} (\bibinfo {year}
  {2000})}\BibitemShut {NoStop}%
\bibitem [{\citenamefont {Acciarri}\ \emph
  {et~al.}(2020{\natexlab{b}})\citenamefont {Acciarri} \emph
  {et~al.}}]{lariat_detpaper}%
  \BibitemOpen
  \bibfield  {author} {\bibinfo {author} {\bibfnamefont {R.}~\bibnamefont
  {Acciarri}} \emph {et~al.} (\bibinfo {collaboration} {LArIAT}),\ }\bibfield
  {title} {\bibinfo {title} {{The Liquid Argon In A Testbeam (LArIAT)
  Experiment}},\ }\href {https://doi.org/10.1088/1748-0221/15/04/P04026}
  {\bibfield  {journal} {\bibinfo  {journal} {{J. Instrum.}}\ }\textbf
  {\bibinfo {volume} {15}},\ \bibinfo {pages} {P04026} (\bibinfo {year}
  {2020}{\natexlab{b}})}\BibitemShut {NoStop}%
\bibitem [{\citenamefont {Amoruso}\ \emph
  {et~al.}(2004{\natexlab{a}})\citenamefont {Amoruso} \emph
  {et~al.}}]{icarus_michel}%
  \BibitemOpen
  \bibfield  {author} {\bibinfo {author} {\bibfnamefont {S.}~\bibnamefont
  {Amoruso}} \emph {et~al.} (\bibinfo {collaboration} {ICARUS}),\ }\bibfield
  {title} {\bibinfo {title} {{Measurement of the $\mu$ decay spectrum with the
  ICARUS liquid argon TPC}},\ }\href
  {https://doi.org/10.1140/epjc/s2004-01597-7} {\bibfield  {journal} {\bibinfo
  {journal} {Eur. Phys. J. C}\ }\textbf {\bibinfo {volume} {33}},\ \bibinfo
  {pages} {233} (\bibinfo {year} {2004}{\natexlab{a}})}\BibitemShut {NoStop}%
\bibitem [{\citenamefont {Acciarri}\ \emph
  {et~al.}(2017{\natexlab{b}})\citenamefont {Acciarri} \emph
  {et~al.}}]{ub_michel}%
  \BibitemOpen
  \bibfield  {author} {\bibinfo {author} {\bibfnamefont {R.}~\bibnamefont
  {Acciarri}} \emph {et~al.} (\bibinfo {collaboration} {MicroBooNE}),\
  }\bibfield  {title} {\bibinfo {title} {{Michel electron reconstruction using
  cosmic-ray data from the MicroBooNE LArTPC}},\ }\href
  {https://doi.org/10.1088/1748-0221/12/09/P09014} {\bibfield  {journal}
  {\bibinfo  {journal} {{J. Instrum.}}\ }\textbf {\bibinfo {volume} {12}},\
  \bibinfo {pages} {P09014} (\bibinfo {year} {2017}{\natexlab{b}})}\BibitemShut
  {NoStop}%
\bibitem [{\citenamefont {Foreman}\ \emph {et~al.}(2020)\citenamefont {Foreman}
  \emph {et~al.}}]{lariat_michels}%
  \BibitemOpen
  \bibfield  {author} {\bibinfo {author} {\bibfnamefont {W.}~\bibnamefont
  {Foreman}} \emph {et~al.} (\bibinfo {collaboration} {LArIAT}),\ }\bibfield
  {title} {\bibinfo {title} {{Calorimetry for low-energy electrons using charge
  and light in liquid argon}},\ }\href
  {https://doi.org/10.1103/PhysRevD.101.012010} {\bibfield  {journal} {\bibinfo
   {journal} {Phys. Rev. D}\ }\textbf {\bibinfo {volume} {101}},\ \bibinfo
  {pages} {012010} (\bibinfo {year} {2020})}\BibitemShut {NoStop}%
\bibitem [{\citenamefont {Amaudruz}\ \emph {et~al.}(2018)\citenamefont
  {Amaudruz} \emph {et~al.}}]{DEAP-3600:2017uua}%
  \BibitemOpen
  \bibfield  {author} {\bibinfo {author} {\bibfnamefont {P.~A.}\ \bibnamefont
  {Amaudruz}} \emph {et~al.} (\bibinfo {collaboration} {DEAP-3600}),\
  }\bibfield  {title} {\bibinfo {title} {{First results from the DEAP-3600 dark
  matter search with argon at SNOLAB}},\ }\href
  {https://doi.org/10.1103/PhysRevLett.121.071801} {\bibfield  {journal}
  {\bibinfo  {journal} {Phys. Rev. Lett.}\ }\textbf {\bibinfo {volume} {121}},\
  \bibinfo {pages} {071801} (\bibinfo {year} {2018})}\BibitemShut {NoStop}%
\bibitem [{\citenamefont {Agnes}\ \emph
  {et~al.}(2018{\natexlab{a}})\citenamefont {Agnes} \emph
  {et~al.}}]{DarkSide:2018kuk}%
  \BibitemOpen
  \bibfield  {author} {\bibinfo {author} {\bibfnamefont {P.}~\bibnamefont
  {Agnes}} \emph {et~al.} (\bibinfo {collaboration} {DarkSide}),\ }\bibfield
  {title} {\bibinfo {title} {{DarkSide-50 532-day Dark Matter Search with
  Low-Radioactivity Argon}},\ }\href
  {https://doi.org/10.1103/PhysRevD.98.102006} {\bibfield  {journal} {\bibinfo
  {journal} {Phys. Rev. D}\ }\textbf {\bibinfo {volume} {98}},\ \bibinfo
  {pages} {102006} (\bibinfo {year} {2018}{\natexlab{a}})}\BibitemShut
  {NoStop}%
\bibitem [{\citenamefont {Agnes}\ \emph
  {et~al.}(2018{\natexlab{b}})\citenamefont {Agnes} \emph
  {et~al.}}]{DarkSide:2018ppu}%
  \BibitemOpen
  \bibfield  {author} {\bibinfo {author} {\bibfnamefont {P.}~\bibnamefont
  {Agnes}} \emph {et~al.} (\bibinfo {collaboration} {DarkSide}),\ }\bibfield
  {title} {\bibinfo {title} {{Constraints on Sub-GeV
  Dark-Matter\textendash{}Electron Scattering from the DarkSide-50
  Experiment}},\ }\href {https://doi.org/10.1103/PhysRevLett.121.111303}
  {\bibfield  {journal} {\bibinfo  {journal} {Phys. Rev. Lett.}\ }\textbf
  {\bibinfo {volume} {121}},\ \bibinfo {pages} {111303} (\bibinfo {year}
  {2018}{\natexlab{b}})}\BibitemShut {NoStop}%
\bibitem [{\citenamefont {Agnes}\ \emph
  {et~al.}(2018{\natexlab{c}})\citenamefont {Agnes} \emph
  {et~al.}}]{DarkSide:2018bpj}%
  \BibitemOpen
  \bibfield  {author} {\bibinfo {author} {\bibfnamefont {P.}~\bibnamefont
  {Agnes}} \emph {et~al.} (\bibinfo {collaboration} {DarkSide}),\ }\bibfield
  {title} {\bibinfo {title} {{Low-Mass Dark Matter Search with the DarkSide-50
  Experiment}},\ }\href {https://doi.org/10.1103/PhysRevLett.121.081307}
  {\bibfield  {journal} {\bibinfo  {journal} {Phys. Rev. Lett.}\ }\textbf
  {\bibinfo {volume} {121}},\ \bibinfo {pages} {081307} (\bibinfo {year}
  {2018}{\natexlab{c}})}\BibitemShut {NoStop}%
\bibitem [{\citenamefont {Castiglioni}\ \emph {et~al.}(2020)\citenamefont
  {Castiglioni}, \citenamefont {Foreman}, \citenamefont {Lepetic},
  \citenamefont {Littlejohn}, \citenamefont {Malaker},\ and\ \citenamefont
  {Mastbaum}}]{Castiglioni:2020tsu}%
  \BibitemOpen
  \bibfield  {author} {\bibinfo {author} {\bibfnamefont {W.}~\bibnamefont
  {Castiglioni}}, \bibinfo {author} {\bibfnamefont {W.}~\bibnamefont
  {Foreman}}, \bibinfo {author} {\bibfnamefont {I.}~\bibnamefont {Lepetic}},
  \bibinfo {author} {\bibfnamefont {B.~R.}\ \bibnamefont {Littlejohn}},
  \bibinfo {author} {\bibfnamefont {M.}~\bibnamefont {Malaker}},\ and\ \bibinfo
  {author} {\bibfnamefont {A.}~\bibnamefont {Mastbaum}},\ }\bibfield  {title}
  {\bibinfo {title} {{Benefits of MeV-scale reconstruction capabilities in
  large liquid argon time projection chambers}},\ }\href
  {https://doi.org/10.1103/PhysRevD.102.092010} {\bibfield  {journal} {\bibinfo
   {journal} {Phys. Rev. D}\ }\textbf {\bibinfo {volume} {102}},\ \bibinfo
  {pages} {092010} (\bibinfo {year} {2020})}\BibitemShut {NoStop}%
\bibitem [{\citenamefont {Andringa}\ \emph {et~al.}(2023)\citenamefont
  {Andringa} \emph {et~al.}}]{leplar_paper}%
  \BibitemOpen
  \bibfield  {author} {\bibinfo {author} {\bibfnamefont {S.}~\bibnamefont
  {Andringa}} \emph {et~al.},\ }\bibfield  {title} {\bibinfo {title}
  {{Low-energy physics in neutrino LArTPCs}},\ }\href
  {https://doi.org/10.1088/1361-6471/acad17} {\bibfield  {journal} {\bibinfo
  {journal} {J. Phys. G}\ }\textbf {\bibinfo {volume} {50}},\ \bibinfo {pages}
  {033001} (\bibinfo {year} {2023})}\BibitemShut {NoStop}%
\bibitem [{\citenamefont {Kubota}\ \emph {et~al.}(2022)\citenamefont {Kubota}
  \emph {et~al.}}]{Q-Pix:2022zjm}%
  \BibitemOpen
  \bibfield  {author} {\bibinfo {author} {\bibfnamefont {S.}~\bibnamefont
  {Kubota}} \emph {et~al.} (\bibinfo {collaboration} {Q-Pix}),\ }\bibfield
  {title} {\bibinfo {title} {{Enhanced low-energy supernova burst detection in
  large liquid argon time projection chambers enabled by Q-Pix}},\ }\href
  {https://doi.org/10.1103/PhysRevD.106.032011} {\bibfield  {journal} {\bibinfo
   {journal} {Phys. Rev. D}\ }\textbf {\bibinfo {volume} {106}},\ \bibinfo
  {pages} {032011} (\bibinfo {year} {2022})}\BibitemShut {NoStop}%
\bibitem [{\citenamefont {Abi}\ \emph {et~al.}(2021)\citenamefont {Abi} \emph
  {et~al.}}]{DUNE:2020zfm}%
  \BibitemOpen
  \bibfield  {author} {\bibinfo {author} {\bibfnamefont {B.}~\bibnamefont
  {Abi}} \emph {et~al.} (\bibinfo {collaboration} {DUNE}),\ }\bibfield  {title}
  {\bibinfo {title} {{Supernova neutrino burst detection with the Deep
  Underground Neutrino Experiment}},\ }\href
  {https://doi.org/10.1140/epjc/s10052-021-09166-w} {\bibfield  {journal}
  {\bibinfo  {journal} {Eur. Phys. J. C}\ }\textbf {\bibinfo {volume} {81}},\
  \bibinfo {pages} {423} (\bibinfo {year} {2021})}\BibitemShut {NoStop}%
\bibitem [{\citenamefont {Capozzi}\ \emph {et~al.}(2019)\citenamefont
  {Capozzi}, \citenamefont {Li}, \citenamefont {Zhu},\ and\ \citenamefont
  {Beacom}}]{dune_solar}%
  \BibitemOpen
  \bibfield  {author} {\bibinfo {author} {\bibfnamefont {F.}~\bibnamefont
  {Capozzi}}, \bibinfo {author} {\bibfnamefont {S.~W.}\ \bibnamefont {Li}},
  \bibinfo {author} {\bibfnamefont {G.}~\bibnamefont {Zhu}},\ and\ \bibinfo
  {author} {\bibfnamefont {J.~F.}\ \bibnamefont {Beacom}},\ }\bibfield  {title}
  {\bibinfo {title} {{DUNE as the Next-Generation Solar Neutrino Experiment}},\
  }\href {https://doi.org/10.1103/PhysRevLett.123.131803} {\bibfield  {journal}
  {\bibinfo  {journal} {Phys. Rev. Lett.}\ }\textbf {\bibinfo {volume} {123}},\
  \bibinfo {pages} {131803} (\bibinfo {year} {2019})}\BibitemShut {NoStop}%
\bibitem [{\citenamefont {Parsa}\ \emph {et~al.}(2022)\citenamefont {Parsa}
  \emph {et~al.}}]{Parsa:2022mnj}%
  \BibitemOpen
  \bibfield  {author} {\bibinfo {author} {\bibfnamefont {S.}~\bibnamefont
  {Parsa}} \emph {et~al.},\ }\bibfield  {title} {\bibinfo {title} {{SoLAr:
  Solar Neutrinos in Liquid Argon}},\ }\Eprint
  {https://arxiv.org/abs/2203.07501} {arXiv:2203.07501 [hep-ex]}  (\bibinfo
  {year} {2022}),\ \bibinfo {note} {{Contribution to Snowmass
  2022}}\BibitemShut {NoStop}%
\bibitem [{\citenamefont {Mastbaum}\ \emph {et~al.}(2022)\citenamefont
  {Mastbaum}, \citenamefont {Psihas},\ and\ \citenamefont
  {Zennamo}}]{Mastbaum:2022rhw}%
  \BibitemOpen
  \bibfield  {author} {\bibinfo {author} {\bibfnamefont {A.}~\bibnamefont
  {Mastbaum}}, \bibinfo {author} {\bibfnamefont {F.}~\bibnamefont {Psihas}},\
  and\ \bibinfo {author} {\bibfnamefont {J.}~\bibnamefont {Zennamo}},\
  }\bibfield  {title} {\bibinfo {title} {{Xenon-doped liquid argon TPCs as a
  neutrinoless double beta decay platform}},\ }\href
  {https://doi.org/10.1103/PhysRevD.106.092002} {\bibfield  {journal} {\bibinfo
   {journal} {Phys. Rev. D}\ }\textbf {\bibinfo {volume} {106}},\ \bibinfo
  {pages} {092002} (\bibinfo {year} {2022})}\BibitemShut {NoStop}%
\bibitem [{\citenamefont {Avasthi}\ \emph {et~al.}(2022)\citenamefont {Avasthi}
  \emph {et~al.}}]{Avasthi:2022tjr}%
  \BibitemOpen
  \bibfield  {author} {\bibinfo {author} {\bibfnamefont {A.}~\bibnamefont
  {Avasthi}} \emph {et~al.},\ }\bibfield  {title} {\bibinfo {title} {{Low
  Background kTon-Scale Liquid Argon Time Projection Chambers}},\ }\Eprint
  {https://arxiv.org/abs/2203.08821} {arXiv:2203.08821 [physics.ins-det]}
  (\bibinfo {year} {2022}),\ \bibinfo {note} {{Contribution to Snowmass
  2021}}\BibitemShut {NoStop}%
\bibitem [{\citenamefont {Abgrall}\ \emph {et~al.}(2021)\citenamefont {Abgrall}
  \emph {et~al.}}]{LEGEND:2021bnm}%
  \BibitemOpen
  \bibfield  {author} {\bibinfo {author} {\bibfnamefont {N.}~\bibnamefont
  {Abgrall}} \emph {et~al.} (\bibinfo {collaboration} {LEGEND}),\ }\bibfield
  {title} {\bibinfo {title} {{The Large Enriched Germanium Experiment for
  Neutrinoless $\beta\beta$ Decay}: {LEGEND-1000 Preconceptual Design
  Report}},\ }\Eprint {https://arxiv.org/abs/2107.11462} {arXiv:2107.11462
  [physics.ins-det]}  (\bibinfo {year} {2021})\BibitemShut {NoStop}%
\bibitem [{\citenamefont {Aalseth}\ \emph {et~al.}(2018)\citenamefont {Aalseth}
  \emph {et~al.}}]{DarkSide-20k:2017zyg}%
  \BibitemOpen
  \bibfield  {author} {\bibinfo {author} {\bibfnamefont {C.~E.}\ \bibnamefont
  {Aalseth}} \emph {et~al.} (\bibinfo {collaboration} {DarkSide-20k}),\
  }\bibfield  {title} {\bibinfo {title} {{DarkSide-20k: A 20 tonne two-phase
  LAr TPC for direct dark matter detection at LNGS}},\ }\href
  {https://doi.org/10.1140/epjp/i2018-11973-4} {\bibfield  {journal} {\bibinfo
  {journal} {Eur. Phys. J. Plus}\ }\textbf {\bibinfo {volume} {133}},\ \bibinfo
  {pages} {131} (\bibinfo {year} {2018})}\BibitemShut {NoStop}%
\bibitem [{\citenamefont {Akerib}\ \emph {et~al.}(2022)\citenamefont {Akerib}
  \emph {et~al.}}]{Akerib:2022ort}%
  \BibitemOpen
  \bibfield  {author} {\bibinfo {author} {\bibfnamefont {D.~S.}\ \bibnamefont
  {Akerib}} \emph {et~al.},\ }\bibfield  {title} {\bibinfo {title}
  {{Snowmass2021 Cosmic Frontier Dark Matter Direct Detection to the Neutrino
  Fog}},\ }\Eprint {https://arxiv.org/abs/2203.08084} {arXiv:2203.08084
  [hep-ex]}  (\bibinfo {year} {2022}),\ \bibinfo {note} {{Contribution to
  Snowmass 2021}}\BibitemShut {NoStop}%
\bibitem [{\citenamefont {Aprile}\ \emph
  {et~al.}(2017{\natexlab{a}})\citenamefont {Aprile} \emph
  {et~al.}}]{XENON:2017fdb}%
  \BibitemOpen
  \bibfield  {author} {\bibinfo {author} {\bibfnamefont {E.}~\bibnamefont
  {Aprile}} \emph {et~al.} (\bibinfo {collaboration} {XENON}),\ }\bibfield
  {title} {\bibinfo {title} {{Material radioassay and selection for the XENON1T
  dark matter experiment}},\ }\href
  {https://doi.org/10.1140/epjc/s10052-017-5329-0} {\bibfield  {journal}
  {\bibinfo  {journal} {Eur. Phys. J. C}\ }\textbf {\bibinfo {volume} {77}},\
  \bibinfo {pages} {890} (\bibinfo {year} {2017}{\natexlab{a}})}\BibitemShut
  {NoStop}%
\bibitem [{\citenamefont {Aprile}\ \emph {et~al.}(2021)\citenamefont {Aprile}
  \emph {et~al.}}]{XENON:2020fbs}%
  \BibitemOpen
  \bibfield  {author} {\bibinfo {author} {\bibfnamefont {E.}~\bibnamefont
  {Aprile}} \emph {et~al.} (\bibinfo {collaboration} {XENON}),\ }\bibfield
  {title} {\bibinfo {title} {{$^{222}$Rn emanation measurements for the XENON1T
  experiment}},\ }\href {https://doi.org/10.1140/epjc/s10052-020-08777-z}
  {\bibfield  {journal} {\bibinfo  {journal} {Eur. Phys. J. C}\ }\textbf
  {\bibinfo {volume} {81}},\ \bibinfo {pages} {337} (\bibinfo {year}
  {2021})}\BibitemShut {NoStop}%
\bibitem [{\citenamefont {Aprile}\ \emph {et~al.}(2022)\citenamefont {Aprile}
  \emph {et~al.}}]{XENON:2021mrg}%
  \BibitemOpen
  \bibfield  {author} {\bibinfo {author} {\bibfnamefont {E.}~\bibnamefont
  {Aprile}} \emph {et~al.} (\bibinfo {collaboration} {XENON}),\ }\bibfield
  {title} {\bibinfo {title} {{Material radiopurity control in the XENONnT
  experiment}},\ }\href {https://doi.org/10.1140/epjc/s10052-022-10345-6}
  {\bibfield  {journal} {\bibinfo  {journal} {Eur. Phys. J. C}\ }\textbf
  {\bibinfo {volume} {82}},\ \bibinfo {pages} {599} (\bibinfo {year}
  {2022})}\BibitemShut {NoStop}%
\bibitem [{\citenamefont {{D. S. Akerib and
  others}}(2020)}]{luxzeplin_cleanliness}%
  \BibitemOpen
  \bibfield  {author} {\bibinfo {author} {\bibnamefont {{D. S. Akerib and
  others}}},\ }\bibfield  {title} {\bibinfo {title} {{The LUX-ZEPLIN (LZ)
  radioactivity and cleanliness control programs}},\ }\href
  {https://doi.org/10.1140/epjc/s10052-020-8420-x} {\bibfield  {journal}
  {\bibinfo  {journal} {Eur. Phys. J.}\ }\textbf {\bibinfo {volume} {80}},\
  \bibinfo {pages} {1044} (\bibinfo {year} {2020})}\BibitemShut {NoStop}%
\bibitem [{\citenamefont {Ackerman}\ \emph {et~al.}(2022)\citenamefont
  {Ackerman} \emph {et~al.}}]{EXO-200:2021srn}%
  \BibitemOpen
  \bibfield  {author} {\bibinfo {author} {\bibfnamefont {N.}~\bibnamefont
  {Ackerman}} \emph {et~al.} (\bibinfo {collaboration} {EXO-200}),\ }\bibfield
  {title} {\bibinfo {title} {{The EXO-200 detector, part~II: auxiliary
  systems}},\ }\href {https://doi.org/10.1088/1748-0221/17/02/P02015}
  {\bibfield  {journal} {\bibinfo  {journal} {{J. Instrum.}}\ }\textbf
  {\bibinfo {volume} {17}},\ \bibinfo {pages} {P02015} (\bibinfo {year}
  {2022})}\BibitemShut {NoStop}%
\bibitem [{\citenamefont {{ Zhicheng Qian and
  others}}(2022)}]{panda_screening}%
  \BibitemOpen
  \bibfield  {author} {\bibinfo {author} {\bibnamefont {{ Zhicheng Qian and
  others}}} (\bibinfo {collaboration} {{PandaX-4T}}),\ }\bibfield  {title}
  {\bibinfo {title} {{Low radioactive material screening and background control
  for the PandaX-4T experiment}},\ }\href
  {https://doi.org/10.1007/JHEP06%282022%29147} {\bibfield  {journal} {\bibinfo
   {journal} {J. High Energ. Phys.}\ }\textbf {\bibinfo {volume} {2022}},\
  \bibinfo {pages} {147}}\BibitemShut {NoStop}%
\bibitem [{\citenamefont {{M. Murra and D. Schulte and C. Huhmann and C.
  Weinheimer}}(2022)}]{XENONnT_rnRemoval}%
  \BibitemOpen
  \bibfield  {author} {\bibinfo {author} {\bibnamefont {{M. Murra and D.
  Schulte and C. Huhmann and C. Weinheimer}}},\ }\bibfield  {title} {\bibinfo
  {title} {{Design, construction and commissioning of a high-flow radon removal
  system for XENONnT}},\ }\href
  {https://link.springer.com/article/10.1140/epjc/s10052-022-11001-9}
  {\bibfield  {journal} {\bibinfo  {journal} {Eur. Phys. J. C}\ }\textbf
  {\bibinfo {volume} {82}},\ \bibinfo {pages} {1104} (\bibinfo {year}
  {2022})}\BibitemShut {NoStop}%
\bibitem [{\citenamefont {Abe}\ \emph {et~al.}(2012{\natexlab{a}})\citenamefont
  {Abe} \emph {et~al.}}]{ABE201250}%
  \BibitemOpen
  \bibfield  {author} {\bibinfo {author} {\bibfnamefont {K.}~\bibnamefont
  {Abe}} \emph {et~al.},\ }\bibfield  {title} {\bibinfo {title} {Radon removal
  from gaseous xenon with activated charcoal},\ }\href
  {https://doi.org/https://doi.org/10.1016/j.nima.2011.09.051} {\bibfield
  {journal} {\bibinfo  {journal} {Nucl. Inst. Meth. A}\ }\textbf {\bibinfo
  {volume} {661}},\ \bibinfo {pages} {50} (\bibinfo {year}
  {2012}{\natexlab{a}})}\BibitemShut {NoStop}%
\bibitem [{\citenamefont {Agnes}\ \emph {et~al.}(2015)\citenamefont {Agnes}
  \emph {et~al.}}]{DarkSide:2014llq}%
  \BibitemOpen
  \bibfield  {author} {\bibinfo {author} {\bibfnamefont {P.}~\bibnamefont
  {Agnes}} \emph {et~al.} (\bibinfo {collaboration} {DarkSide}),\ }\bibfield
  {title} {\bibinfo {title} {{First Results from the DarkSide-50 Dark Matter
  Experiment at Laboratori Nazionali del Gran Sasso}},\ }\href
  {https://doi.org/10.1016/j.physletb.2015.03.012} {\bibfield  {journal}
  {\bibinfo  {journal} {Phys. Lett. B}\ }\textbf {\bibinfo {volume} {743}},\
  \bibinfo {pages} {456} (\bibinfo {year} {2015})}\BibitemShut {NoStop}%
\bibitem [{\citenamefont {Aprile}\ \emph
  {et~al.}(2017{\natexlab{b}})\citenamefont {Aprile} \emph
  {et~al.}}]{XENON100:2017gsw}%
  \BibitemOpen
  \bibfield  {author} {\bibinfo {author} {\bibfnamefont {E.}~\bibnamefont
  {Aprile}} \emph {et~al.} (\bibinfo {collaboration} {XENON100}),\ }\bibfield
  {title} {\bibinfo {title} {{Online $^{222}$Rn removal by cryogenic
  distillation in the XENON100 experiment}},\ }\href
  {https://doi.org/10.1140/epjc/s10052-017-4902-x} {\bibfield  {journal}
  {\bibinfo  {journal} {Eur. Phys. J. C}\ }\textbf {\bibinfo {volume} {77}},\
  \bibinfo {pages} {358} (\bibinfo {year} {2017}{\natexlab{b}})}\BibitemShut
  {NoStop}%
\bibitem [{\citenamefont {Abe}\ \emph {et~al.}(2012{\natexlab{b}})\citenamefont
  {Abe} \emph {et~al.}}]{gas_xe_rn}%
  \BibitemOpen
  \bibfield  {author} {\bibinfo {author} {\bibfnamefont {K.}~\bibnamefont
  {Abe}} \emph {et~al.},\ }\bibfield  {title} {\bibinfo {title} {{Radon removal
  from gaseous xenon with activated charcoal}},\ }\href
  {https://doi.org/https://doi.org/10.1016/j.nima.2011.09.051} {\bibfield
  {journal} {\bibinfo  {journal} {Nucl. Instrum. Meth. A}\ }\textbf {\bibinfo
  {volume} {661}},\ \bibinfo {pages} {50} (\bibinfo {year}
  {2012}{\natexlab{b}})}\BibitemShut {NoStop}%
\bibitem [{\citenamefont {{E. Aprile and others}}(2017)}]{xenon100_rn_removal}%
  \BibitemOpen
  \bibfield  {author} {\bibinfo {author} {\bibnamefont {{E. Aprile and
  others}}} (\bibinfo {collaboration} {{XENON}}),\ }\bibfield  {title}
  {\bibinfo {title} {{Online 222Rn Removal by cryogenic distallation in the
  XENON100 experiment}},\ }\href
  {https://link.springer.com/article/10.1140/epjc/s10052-017-4902-x} {\bibfield
   {journal} {\bibinfo  {journal} {Eur. Phys. J. C}\ }\textbf {\bibinfo
  {volume} {77}},\ \bibinfo {pages} {358} (\bibinfo {year} {2017})}\BibitemShut
  {NoStop}%
\bibitem [{\citenamefont {Ajaj}\ \emph {et~al.}(2019)\citenamefont {Ajaj} \emph
  {et~al.}}]{PhysRevD.100.022004}%
  \BibitemOpen
  \bibfield  {author} {\bibinfo {author} {\bibfnamefont {R.}~\bibnamefont
  {Ajaj}} \emph {et~al.} (\bibinfo {collaboration} {DEAP}),\ }\bibfield
  {title} {\bibinfo {title} {{Search for dark matter with a 231-day exposure of
  liquid argon using DEAP-3600 at SNOLAB}},\ }\href
  {https://doi.org/10.1103/PhysRevD.100.022004} {\bibfield  {journal} {\bibinfo
   {journal} {Phys. Rev. D}\ }\textbf {\bibinfo {volume} {100}},\ \bibinfo
  {pages} {022004} (\bibinfo {year} {2019})}\BibitemShut {NoStop}%
\bibitem [{\citenamefont {Adamowski}\ \emph {et~al.}(2014)\citenamefont
  {Adamowski} \emph {et~al.}}]{Adamowski:2014daa}%
  \BibitemOpen
  \bibfield  {author} {\bibinfo {author} {\bibfnamefont {M.}~\bibnamefont
  {Adamowski}} \emph {et~al.},\ }\bibfield  {title} {\bibinfo {title} {{The
  Liquid Argon Purity Demonstrator}},\ }\href
  {https://doi.org/10.1088/1748-0221/9/07/P07005} {\bibfield  {journal}
  {\bibinfo  {journal} {{J. Instrum.}}\ }\textbf {\bibinfo {volume} {9}},\
  \bibinfo {pages} {P07005} (\bibinfo {year} {2014})}\BibitemShut {NoStop}%
\bibitem [{\citenamefont {Andrews}\ \emph {et~al.}(2009)\citenamefont {Andrews}
  \emph {et~al.}}]{Andrews:2009zza}%
  \BibitemOpen
  \bibfield  {author} {\bibinfo {author} {\bibfnamefont {R.}~\bibnamefont
  {Andrews}} \emph {et~al.},\ }\bibfield  {title} {\bibinfo {title} {{A system
  to test the effects of materials on the electron drift lifetime in liquid
  argon and observations on the effect of water}},\ }\href
  {https://doi.org/10.1016/j.nima.2009.07.024} {\bibfield  {journal} {\bibinfo
  {journal} {Nucl. Instrum. Meth. A}\ }\textbf {\bibinfo {volume} {608}},\
  \bibinfo {pages} {251} (\bibinfo {year} {2009})}\BibitemShut {NoStop}%
\bibitem [{\citenamefont {Reichenbacher}(2020)}]{JuergenTalk2}%
  \BibitemOpen
  \bibfield  {author} {\bibinfo {author} {\bibfnamefont {J.}~\bibnamefont
  {Reichenbacher}},\ }\href
  {https://indico.fnal.gov/event/43870/contributions/188848/attachments/131670/161298/DUNE_BackgroundsWorkshop_radonOverview_20July2020_JR.pdf}
  {\bibinfo {title} {{Radon-Induced Backgrounds: Introduction with Sources and
  Assay Overview}}},\ \bibinfo {howpublished} {{DUNE Background Mitigation
  Strategies Workshop}} (\bibinfo {year} {2020})\BibitemShut {NoStop}%
\bibitem [{\citenamefont {Abratenko}(2022)}]{ub_radon}%
  \BibitemOpen
  \bibfield  {author} {\bibinfo {author} {\bibfnamefont {P.}~\bibnamefont
  {Abratenko}} (\bibinfo {collaboration} {MicroBooNE}),\ }\bibfield  {title}
  {\bibinfo {title} {{Observation of radon mitigation in MicroBooNE by a liquid
  argon filtration system}},\ }\href
  {https://doi.org/10.1088/1748-0221/17/11/P11022} {\bibfield  {journal}
  {\bibinfo  {journal} {{J. Instrum.}}\ }\textbf {\bibinfo {volume} {17}},\
  \bibinfo {pages} {P11022} (\bibinfo {year} {2022})}\BibitemShut {NoStop}%
\bibitem [{\citenamefont {{Sigma-Aldrich, P.O. Box 14508, St. Louis, MO 63178
  USA}}()}]{filter_ref_sieve}%
  \BibitemOpen
  \bibfield  {author} {\bibinfo {author} {\bibnamefont {{Sigma-Aldrich, P.O.
  Box 14508, St. Louis, MO 63178 USA}}},\ }\href@noop {} {}\bibinfo {note}
  {\href{https://www.sigmaaldrich.com/US/en/product/sigald/208604}{\nolinkurl{https://www.sigmaaldrich.com/US/en/product/sigald/208604}}}\BibitemShut
  {NoStop}%
\bibitem [{\citenamefont {{BASF Corp., 100 Park Avenue, Florham Park, NJ 07932
  USA}}()}]{filter_ref_pellets}%
  \BibitemOpen
  \bibfield  {author} {\bibinfo {author} {\bibnamefont {{BASF Corp., 100 Park
  Avenue, Florham Park, NJ 07932 USA}}},\ }\href@noop {} {}\bibinfo {note}
  {\href{https://catalysts.basf.com/products/cu-0226-s-1}{\nolinkurl{https://catalysts.basf.com/products/cu-0226-s-1}}}\BibitemShut
  {NoStop}%
\bibitem [{\citenamefont {Acciarri}\ \emph
  {et~al.}(2017{\natexlab{c}})\citenamefont {Acciarri} \emph
  {et~al.}}]{ub_noise}%
  \BibitemOpen
  \bibfield  {author} {\bibinfo {author} {\bibfnamefont {R.}~\bibnamefont
  {Acciarri}} \emph {et~al.} (\bibinfo {collaboration} {MicroBooNE}),\
  }\bibfield  {title} {\bibinfo {title} {{Noise Characterization and Filtering
  in the MicroBooNE Liquid Argon TPC}},\ }\href
  {https://doi.org/10.1088/1748-0221/12/08/P08003} {\bibfield  {journal}
  {\bibinfo  {journal} {{J. Instrum.}}\ }\textbf {\bibinfo {volume} {12}},\
  \bibinfo {pages} {P08003} (\bibinfo {year} {2017}{\natexlab{c}})}\BibitemShut
  {NoStop}%
\bibitem [{\citenamefont {Amaudruz}\ \emph {et~al.}(2015)\citenamefont
  {Amaudruz} \emph {et~al.}}]{Amaudruz:2012hr}%
  \BibitemOpen
  \bibfield  {author} {\bibinfo {author} {\bibfnamefont {P.~A.}\ \bibnamefont
  {Amaudruz}} \emph {et~al.},\ }\bibfield  {title} {\bibinfo {title} {{Radon
  backgrounds in the DEAP-1 liquid argon based Dark Matter detector}},\ }\href
  {https://doi.org/10.1016/j.astropartphys.2014.09.006} {\bibfield  {journal}
  {\bibinfo  {journal} {Astropart. Phys.}\ }\textbf {\bibinfo {volume} {62}},\
  \bibinfo {pages} {178} (\bibinfo {year} {2015})}\BibitemShut {NoStop}%
\bibitem [{\citenamefont {Adams}\ \emph
  {et~al.}(2020{\natexlab{a}})\citenamefont {Adams} \emph {et~al.}}]{ub_cal}%
  \BibitemOpen
  \bibfield  {author} {\bibinfo {author} {\bibfnamefont {C.}~\bibnamefont
  {Adams}} \emph {et~al.} (\bibinfo {collaboration} {MicroBooNE}),\ }\bibfield
  {title} {\bibinfo {title} {{Calibration of the charge and energy loss per
  unit length of the MicroBooNE liquid argon time projection chamber using
  muons and protons}},\ }\href {https://doi.org/10.1088/1748-0221/15/03/P03022}
  {\bibfield  {journal} {\bibinfo  {journal} {{J. Instrum.}}\ }\textbf
  {\bibinfo {volume} {15}},\ \bibinfo {pages} {P03022} (\bibinfo {year}
  {2020}{\natexlab{a}})}\BibitemShut {NoStop}%
\bibitem [{\citenamefont {Abratenko}\ \emph {et~al.}(2022)\citenamefont
  {Abratenko} \emph {et~al.}}]{ub_syst}%
  \BibitemOpen
  \bibfield  {author} {\bibinfo {author} {\bibfnamefont {P.}~\bibnamefont
  {Abratenko}} \emph {et~al.} (\bibinfo {collaboration} {MicroBooNE}),\
  }\bibfield  {title} {\bibinfo {title} {{Novel approach for evaluating
  detector-related uncertainties in a LArTPC using MicroBooNE data}},\ }\href
  {https://doi.org/10.1140/epjc/s10052-022-10270-8} {\bibfield  {journal}
  {\bibinfo  {journal} {Eur. Phys. J. C}\ }\textbf {\bibinfo {volume} {82}},\
  \bibinfo {pages} {454} (\bibinfo {year} {2022})}\BibitemShut {NoStop}%
\bibitem [{\citenamefont {{Pylon Electronics, Inc.}}()}]{radium_source}%
  \BibitemOpen
  \bibfield  {author} {\bibinfo {author} {\bibnamefont {{Pylon Electronics,
  Inc.}}},\ }\href@noop {} {\bibinfo {title} {{Model 2000A Radioactive
  Source}}},\ \bibinfo {note}
  {\href{https://pylonelectronics-radon.com/radioactive-sources/}{\nolinkurl{https://pylonelectronics-radon.com/radioactive-sources/}}}\BibitemShut
  {NoStop}%
\bibitem [{\citenamefont {Snider}\ and\ \citenamefont
  {Petrillo}(2017)}]{larsoft}%
  \BibitemOpen
  \bibfield  {author} {\bibinfo {author} {\bibfnamefont {E.}~\bibnamefont
  {Snider}}\ and\ \bibinfo {author} {\bibfnamefont {G.}~\bibnamefont
  {Petrillo}},\ }\bibfield  {title} {\bibinfo {title} {{LArSoft: toolkit for
  simulation, reconstruction and analysis of liquid argon TPC neutrino
  detectors}},\ }\href {https://doi.org/10.1088/1742-6596/898/4/042057}
  {\bibfield  {journal} {\bibinfo  {journal} {J. Phys. Conf. Ser.}\ }\textbf
  {\bibinfo {volume} {898}},\ \bibinfo {pages} {042057} (\bibinfo {year}
  {2017})}\BibitemShut {NoStop}%
\bibitem [{\citenamefont {Adams}\ \emph
  {et~al.}(2018{\natexlab{a}})\citenamefont {Adams} \emph
  {et~al.}}]{ub_sigproc_pt1}%
  \BibitemOpen
  \bibfield  {author} {\bibinfo {author} {\bibfnamefont {C.}~\bibnamefont
  {Adams}} \emph {et~al.} (\bibinfo {collaboration} {{MicroBooNE}}),\
  }\bibfield  {title} {\bibinfo {title} {{Ionization electron signal processing
  in single phase LArTPCs. Part I. Algorithm Description and quantitative
  evaluation with MicroBooNE simulation}},\ }\href
  {https://doi.org/10.1088/1748-0221/13/07/P07006} {\bibfield  {journal}
  {\bibinfo  {journal} {{J. Instrum.}}\ }\textbf {\bibinfo {volume} {13}},\
  \bibinfo {pages} {P07006} (\bibinfo {year} {2018}{\natexlab{a}})}\BibitemShut
  {NoStop}%
\bibitem [{\citenamefont {Adams}\ \emph
  {et~al.}(2018{\natexlab{b}})\citenamefont {Adams} \emph
  {et~al.}}]{ub_sigproc_pt2}%
  \BibitemOpen
  \bibfield  {author} {\bibinfo {author} {\bibfnamefont {C.}~\bibnamefont
  {Adams}} \emph {et~al.} (\bibinfo {collaboration} {{MicroBooNE}}),\
  }\bibfield  {title} {\bibinfo {title} {{Ionization electron signal processing
  in single phase LArTPCs. Part II. Data/simulation comparison and performance
  in MicroBooNE}},\ }\href {https://doi.org/10.1088/1748-0221/13/07/P07007}
  {\bibfield  {journal} {\bibinfo  {journal} {{J. Instrum.}}\ }\textbf
  {\bibinfo {volume} {13}},\ \bibinfo {pages} {P07007} (\bibinfo {year}
  {2018}{\natexlab{b}})}\BibitemShut {NoStop}%
\bibitem [{\citenamefont {Baller}(2017)}]{Baller_2017}%
  \BibitemOpen
  \bibfield  {author} {\bibinfo {author} {\bibfnamefont {B.}~\bibnamefont
  {Baller}},\ }\bibfield  {title} {\bibinfo {title} {{Liquid argon TPC signal
  formation, signal processing and reconstruction techniques}},\ }\href
  {https://doi.org/10.1088/1748-0221/12/07/p07010} {\bibfield  {journal}
  {\bibinfo  {journal} {{J. Instrum.}}\ }\textbf {\bibinfo {volume} {12}},\
  \bibinfo {pages} {P07010} (\bibinfo {year} {2017})}\BibitemShut {NoStop}%
\bibitem [{\citenamefont {Acciarri}\ \emph {et~al.}(2018)\citenamefont
  {Acciarri} \emph {et~al.}}]{ub_pandora}%
  \BibitemOpen
  \bibfield  {author} {\bibinfo {author} {\bibfnamefont {R.}~\bibnamefont
  {Acciarri}} \emph {et~al.} (\bibinfo {collaboration} {{MicroBooNE}}),\
  }\bibfield  {title} {\bibinfo {title} {{The Pandora multi-algorithm approach
  to automated pattern recognition of cosmic-ray muon and neutrino events in
  the MicroBooNE detector}},\ }\href
  {https://doi.org/10.1140/epjc/s10052-017-5481-6} {\bibfield  {journal}
  {\bibinfo  {journal} {Eur. Phys. J. C}\ }\textbf {\bibinfo {volume} {78}},\
  \bibinfo {pages} {82} (\bibinfo {year} {2018})}\BibitemShut {NoStop}%
\bibitem [{\citenamefont {Miyajima}\ \emph {et~al.}(1974)\citenamefont
  {Miyajima} \emph {et~al.}}]{PhysRevA.9.1438}%
  \BibitemOpen
  \bibfield  {author} {\bibinfo {author} {\bibfnamefont {M.}~\bibnamefont
  {Miyajima}} \emph {et~al.},\ }\bibfield  {title} {\bibinfo {title} {Average
  energy expended per ion pair in liquid argon},\ }\href
  {https://doi.org/10.1103/PhysRevA.9.1438} {\bibfield  {journal} {\bibinfo
  {journal} {Phys. Rev. A}\ }\textbf {\bibinfo {volume} {9}},\ \bibinfo {pages}
  {1438} (\bibinfo {year} {1974})}\BibitemShut {NoStop}%
\bibitem [{\citenamefont {Abratenko}\ \emph {et~al.}(2020)\citenamefont
  {Abratenko} \emph {et~al.}}]{Abratenko_2020}%
  \BibitemOpen
  \bibfield  {author} {\bibinfo {author} {\bibfnamefont {P.}~\bibnamefont
  {Abratenko}} \emph {et~al.} (\bibinfo {collaboration} {{MicroBooNE}}),\
  }\bibfield  {title} {\bibinfo {title} {Measurement of space charge effects in
  the {MicroBooNE} {LArTPC} using cosmic muons},\ }\href
  {https://doi.org/10.1088/1748-0221/15/12/p12037} {\bibfield  {journal}
  {\bibinfo  {journal} {{J. Instrum.}}\ }\textbf {\bibinfo {volume} {15}},\
  \bibinfo {pages} {P12037} (\bibinfo {year} {2020})}\BibitemShut {NoStop}%
\bibitem [{\citenamefont {{ESTAR: Stopping Power and Range Tables for
  Electrons}}(2020)}]{csda}%
  \BibitemOpen
  \bibfield  {author} {\bibinfo {author} {\bibnamefont {{ESTAR: Stopping Power
  and Range Tables for Electrons}}},\ }\href@noop {} {} (\bibinfo {year}
  {2020}),\ \bibinfo {note}
  {\href{https://physics.nist.gov/PhysRefData/Star/Text/ESTAR.html}{https://physics.nist.gov/}}\BibitemShut
  {NoStop}%
\bibitem [{\citenamefont {Acciarri}\ \emph {et~al.}(2013)\citenamefont
  {Acciarri} \emph {et~al.}}]{argo_modbox}%
  \BibitemOpen
  \bibfield  {author} {\bibinfo {author} {\bibfnamefont {R.}~\bibnamefont
  {Acciarri}} \emph {et~al.} (\bibinfo {collaboration} {ArgoNeuT
  Collaboration}),\ }\bibfield  {title} {\bibinfo {title} {{A study of electron
  recombination using highly ionizing particles in the ArgoNeuT Liquid Argon
  TPC}},\ }\href {https://doi.org/10.1088/1748-0221/8/08/P08005} {\bibfield
  {journal} {\bibinfo  {journal} {{J. Instrum.}}\ }\textbf {\bibinfo {volume}
  {8}},\ \bibinfo {pages} {P08005} (\bibinfo {year} {2013})}\BibitemShut
  {NoStop}%
\bibitem [{\citenamefont {Kolanoski}\ and\ \citenamefont
  {Wermes}(2020)}]{kolanoski_book}%
  \BibitemOpen
  \bibfield  {author} {\bibinfo {author} {\bibfnamefont {H.}~\bibnamefont
  {Kolanoski}}\ and\ \bibinfo {author} {\bibfnamefont {N.}~\bibnamefont
  {Wermes}},\ }\href
  {{https://books.google.com/books?hl=en&lr=&id=2TjpDwAAQBAJ&oi=fnd&pg=PP1&ots=CQAbxiTi6L&sig=8SoAPqvcX0sdYECiCk55Wqq8XaU#v=onepage&q&f=false}}
  {\emph {\bibinfo {title} {{Particle Detectors: Fundamentals and
  Applications}}}}\ (\bibinfo  {publisher} {{Oxford University Press}},\
  \bibinfo {year} {2020})\ p.\ \bibinfo {pages} {623}\BibitemShut {NoStop}%
\bibitem [{\citenamefont {Toth}(1977)}]{TOTH1977437}%
  \BibitemOpen
  \bibfield  {author} {\bibinfo {author} {\bibfnamefont {K.}~\bibnamefont
  {Toth}},\ }\bibfield  {title} {\bibinfo {title} {{Nuclear data sheets for A =
  214}},\ }\href
  {https://doi.org/https://doi.org/10.1016/S0090-3752(77)80026-4} {\bibfield
  {journal} {\bibinfo  {journal} {Nuclear Data Sheets}\ }\textbf {\bibinfo
  {volume} {21}},\ \bibinfo {pages} {437} (\bibinfo {year} {1977})}\BibitemShut
  {NoStop}%
\bibitem [{\citenamefont {Adams}\ \emph
  {et~al.}(2020{\natexlab{b}})\citenamefont {Adams} \emph
  {et~al.}}]{ub_efield}%
  \BibitemOpen
  \bibfield  {author} {\bibinfo {author} {\bibfnamefont {C.}~\bibnamefont
  {Adams}} \emph {et~al.} (\bibinfo {collaboration} {{MicroBooNE}}),\
  }\bibfield  {title} {\bibinfo {title} {{A method to determine the electric
  field of liquid argon time projection chambers using a UV laser system and
  its application in MicroBooNE}},\ }\href
  {https://doi.org/10.1088/1748-0221/15/07/P07010} {\bibfield  {journal}
  {\bibinfo  {journal} {{J. Instrum.}}\ }\textbf {\bibinfo {volume} {15}},\
  \bibinfo {pages} {P07010} (\bibinfo {year} {2020}{\natexlab{b}})}\BibitemShut
  {NoStop}%
\bibitem [{\citenamefont {Hitachi}\ \emph {et~al.}(1987)\citenamefont
  {Hitachi}, \citenamefont {Yunoki}, \citenamefont {Doke},\ and\ \citenamefont
  {Takahashi}}]{PhysRevA.35.3956}%
  \BibitemOpen
  \bibfield  {author} {\bibinfo {author} {\bibfnamefont {A.}~\bibnamefont
  {Hitachi}}, \bibinfo {author} {\bibfnamefont {A.}~\bibnamefont {Yunoki}},
  \bibinfo {author} {\bibfnamefont {T.}~\bibnamefont {Doke}},\ and\ \bibinfo
  {author} {\bibfnamefont {T.}~\bibnamefont {Takahashi}},\ }\bibfield  {title}
  {\bibinfo {title} {{Scintillation and ionization yield for
  \ensuremath{\alpha} particles and fission fragments in liquid argon}},\
  }\href {https://doi.org/10.1103/PhysRevA.35.3956} {\bibfield  {journal}
  {\bibinfo  {journal} {Phys. Rev. A}\ }\textbf {\bibinfo {volume} {35}},\
  \bibinfo {pages} {3956} (\bibinfo {year} {1987})}\BibitemShut {NoStop}%
\bibitem [{\citenamefont {Agnes}\ \emph {et~al.}(2017)\citenamefont {Agnes}
  \emph {et~al.}}]{DarkSide:2016ddo}%
  \BibitemOpen
  \bibfield  {author} {\bibinfo {author} {\bibfnamefont {P.}~\bibnamefont
  {Agnes}} \emph {et~al.} (\bibinfo {collaboration} {DarkSide}),\ }\bibfield
  {title} {\bibinfo {title} {{Effect of Low Electric Fields on Alpha
  Scintillation Light Yield in Liquid Argon}},\ }\href
  {https://doi.org/10.1088/1748-0221/12/01/P01021} {\bibfield  {journal}
  {\bibinfo  {journal} {{J. Instrum.}}\ }\textbf {\bibinfo {volume} {12}},\
  \bibinfo {pages} {P01021} (\bibinfo {year} {2017})}\BibitemShut {NoStop}%
\bibitem [{\citenamefont {{National Nuclear Data Center (NNDC), Brookhaven
  National Laboratory}}(2022)}]{nndc_chart}%
  \BibitemOpen
  \bibfield  {author} {\bibinfo {author} {\bibnamefont {{National Nuclear Data
  Center (NNDC), Brookhaven National Laboratory}}},\ }\href
  {https://www.nndc.bnl.gov/nudat3} {\bibinfo {title} {{NuDat 3.0 Database}}}
  (\bibinfo {year} {2022})\BibitemShut {NoStop}%
\bibitem [{\citenamefont {Ponkratenko}\ \emph {et~al.}(2000)\citenamefont
  {Ponkratenko}, \citenamefont {Tretyak},\ and\ \citenamefont
  {Zdesenko}}]{Ponkratenko:2000um}%
  \BibitemOpen
  \bibfield  {author} {\bibinfo {author} {\bibfnamefont {O.~A.}\ \bibnamefont
  {Ponkratenko}}, \bibinfo {author} {\bibfnamefont {V.~I.}\ \bibnamefont
  {Tretyak}},\ and\ \bibinfo {author} {\bibfnamefont {Y.~G.}\ \bibnamefont
  {Zdesenko}},\ }\bibfield  {title} {\bibinfo {title} {{The Event generator
  DECAY4 for simulation of double beta processes and decay of radioactive
  nuclei}},\ }\href {https://doi.org/10.1134/1.855784} {\bibfield  {journal}
  {\bibinfo  {journal} {Phys. At. Nucl.}\ }\textbf {\bibinfo {volume} {63}},\
  \bibinfo {pages} {1282} (\bibinfo {year} {2000})}\BibitemShut {NoStop}%
\bibitem [{\citenamefont {Agostinelli}\ \emph {et~al.}(2003)\citenamefont
  {Agostinelli} \emph {et~al.}}]{g4}%
  \BibitemOpen
  \bibfield  {author} {\bibinfo {author} {\bibfnamefont {S.}~\bibnamefont
  {Agostinelli}} \emph {et~al.},\ }\bibfield  {title} {\bibinfo {title}
  {{Geant4 - A Simulation Toolkit}},\ }\href
  {https://doi.org/https://doi.org/10.1016/S0168-9002(03)01368-8} {\bibfield
  {journal} {\bibinfo  {journal} {Nucl. Instrum. Meth.}\ }\textbf {\bibinfo
  {volume} {A506}},\ \bibinfo {pages} {250} (\bibinfo {year}
  {2003})}\BibitemShut {NoStop}%
\bibitem [{\citenamefont {Abratenko}\ \emph {et~al.}(2021)\citenamefont
  {Abratenko} \emph {et~al.}}]{ub_diffusion}%
  \BibitemOpen
  \bibfield  {author} {\bibinfo {author} {\bibfnamefont {P.}~\bibnamefont
  {Abratenko}} \emph {et~al.} (\bibinfo {collaboration} {{MicroBooNE}}),\
  }\bibfield  {title} {\bibinfo {title} {{Measurement of the longitudinal
  diffusion of ionization electrons in the MicroBooNE detector}},\ }\href
  {https://doi.org/10.1088/1748-0221/16/09/P09025} {\bibfield  {journal}
  {\bibinfo  {journal} {{J. Instrum.}}\ }\textbf {\bibinfo {volume} {16}},\
  \bibinfo {pages} {P09025} (\bibinfo {year} {2021})}\BibitemShut {NoStop}%
\bibitem [{\citenamefont {Mei}\ \emph {et~al.}(2008)\citenamefont {Mei},
  \citenamefont {Yin}, \citenamefont {Stonehill},\ and\ \citenamefont
  {Hime}}]{MEI200812}%
  \BibitemOpen
  \bibfield  {author} {\bibinfo {author} {\bibfnamefont {D.-M.}\ \bibnamefont
  {Mei}}, \bibinfo {author} {\bibfnamefont {Z.-B.}\ \bibnamefont {Yin}},
  \bibinfo {author} {\bibfnamefont {L.}~\bibnamefont {Stonehill}},\ and\
  \bibinfo {author} {\bibfnamefont {A.}~\bibnamefont {Hime}},\ }\bibfield
  {title} {\bibinfo {title} {A model of nuclear recoil scintillation efficiency
  in noble liquids},\ }\href
  {https://doi.org/https://doi.org/10.1016/j.astropartphys.2008.06.001}
  {\bibfield  {journal} {\bibinfo  {journal} {Astropart. Phys.}\ }\textbf
  {\bibinfo {volume} {30}},\ \bibinfo {pages} {12} (\bibinfo {year}
  {2008})}\BibitemShut {NoStop}%
\bibitem [{\citenamefont {Hitachi}(2005)}]{HITACHI2005247}%
  \BibitemOpen
  \bibfield  {author} {\bibinfo {author} {\bibfnamefont {A.}~\bibnamefont
  {Hitachi}},\ }\bibfield  {title} {\bibinfo {title} {Properties of liquid
  xenon scintillation for dark matter searches},\ }\href
  {https://doi.org/https://doi.org/10.1016/j.astropartphys.2005.07.002}
  {\bibfield  {journal} {\bibinfo  {journal} {Astropart. Phys.}\ }\textbf
  {\bibinfo {volume} {24}},\ \bibinfo {pages} {247} (\bibinfo {year}
  {2005})}\BibitemShut {NoStop}%
\bibitem [{\citenamefont {Hitachi}(2004)}]{Hitachi:2004zw}%
  \BibitemOpen
  \bibfield  {author} {\bibinfo {author} {\bibfnamefont {A.}~\bibnamefont
  {Hitachi}},\ }\bibfield  {title} {\bibinfo {title} {{Properties of liquid
  rare gas scintillation for WIMP searches}},\ }in\ \href
  {https://doi.org/https://doi.org/10.1142/9789812701848_0059} {\emph {\bibinfo
  {booktitle} {{Proceedings of the 5th International Workshop on the
  Identification of Dark Matter}}}}\ (\bibinfo {year} {2004})\ pp.\ \bibinfo
  {pages} {396--401}\BibitemShut {NoStop}%
\bibitem [{\citenamefont {Szydagis}\ \emph {et~al.}(2021)\citenamefont
  {Szydagis} \emph {et~al.}}]{nest_paper}%
  \BibitemOpen
  \bibfield  {author} {\bibinfo {author} {\bibfnamefont {M.}~\bibnamefont
  {Szydagis}} \emph {et~al.},\ }\bibfield  {title} {\bibinfo {title} {{A Review
  of Basic Energy Reconstruction Techniques in Liquid Xenon and Argon Detectors
  for Dark Matter and Neutrino Physics Using NEST}},\ }\href
  {https://www.mdpi.com/2410-390X/5/1/13} {\bibfield  {journal} {\bibinfo
  {journal} {Instruments}\ }\textbf {\bibinfo {volume} {5}},\ \bibinfo {pages}
  {13} (\bibinfo {year} {2021})}\BibitemShut {NoStop}%
\bibitem [{\citenamefont {Szydagis}\ \emph {et~al.}(2023)\citenamefont
  {Szydagis} \emph {et~al.}}]{nest_software}%
  \BibitemOpen
  \bibfield  {author} {\bibinfo {author} {\bibnamefont {Szydagis}} \emph
  {et~al.},\ }\href@noop {} {\bibinfo {title} {{Noble Element Simulation
  Technique (v2.3.12)}}} (\bibinfo {year} {2023}),\ \bibinfo {note}
  {\href{https://doi.org/10.5281/zenodo.7577399}{\nolinkurl{https://doi.org/10.5281/zenodo.7577399}}}\BibitemShut
  {NoStop}%
\bibitem [{\citenamefont {Amoruso}\ \emph
  {et~al.}(2004{\natexlab{b}})\citenamefont {Amoruso} \emph
  {et~al.}}]{lar_diff2}%
  \BibitemOpen
  \bibfield  {author} {\bibinfo {author} {\bibfnamefont {S.}~\bibnamefont
  {Amoruso}} \emph {et~al.} (\bibinfo {collaboration} {{ICARUS}}),\ }\bibfield
  {title} {\bibinfo {title} {{Study of electron recombination in liquid argon
  with the ICARUS TPC}},\ }\href
  {https://doi.org/https://doi.org/10.1016/j.nima.2003.11.423} {\bibfield
  {journal} {\bibinfo  {journal} {Nucl. Instrum. Meth. A}\ }\textbf {\bibinfo
  {volume} {523}},\ \bibinfo {pages} {275 } (\bibinfo {year}
  {2004}{\natexlab{b}})}\BibitemShut {NoStop}%
\bibitem [{\citenamefont {Aprile}\ \emph {et~al.}(1987)\citenamefont {Aprile},
  \citenamefont {Ku}, \citenamefont {Park},\ and\ \citenamefont
  {Schwartz}}]{APRILE1987519}%
  \BibitemOpen
  \bibfield  {author} {\bibinfo {author} {\bibfnamefont {E.}~\bibnamefont
  {Aprile}}, \bibinfo {author} {\bibfnamefont {W.~H.-M.}\ \bibnamefont {Ku}},
  \bibinfo {author} {\bibfnamefont {J.}~\bibnamefont {Park}},\ and\ \bibinfo
  {author} {\bibfnamefont {H.}~\bibnamefont {Schwartz}},\ }\bibfield  {title}
  {\bibinfo {title} {{Energy resolution studies of liquid argon ionization
  detectors}},\ }\href
  {https://doi.org/https://doi.org/10.1016/0168-9002(87)90362-7} {\bibfield
  {journal} {\bibinfo  {journal} {Nucl. Instrum. Meth. A}\ }\textbf {\bibinfo
  {volume} {261}},\ \bibinfo {pages} {519} (\bibinfo {year}
  {1987})}\BibitemShut {NoStop}%
\bibitem [{\citenamefont {Aprile}\ \emph {et~al.}(2018)\citenamefont {Aprile}
  \emph {et~al.}}]{XENON_RnKr}%
  \BibitemOpen
  \bibfield  {author} {\bibinfo {author} {\bibfnamefont {E.}~\bibnamefont
  {Aprile}} \emph {et~al.} (\bibinfo {collaboration} {XENON}),\ }\bibfield
  {title} {\bibinfo {title} {{Intrinsic backgrounds from Rn and Kr in the
  XENON100 experiment}},\ }\href
  {https://doi.org/10.1140/epjc/s10052-018-5565-y} {\bibfield  {journal}
  {\bibinfo  {journal} {Eur. Phys. J. C}\ }\textbf {\bibinfo {volume} {78}},\
  \bibinfo {pages} {132} (\bibinfo {year} {2018})}\BibitemShut {NoStop}%
\bibitem [{\citenamefont {{Aalbers, J. and
  others}}(2023)}]{PhysRevD.108.012010}%
  \BibitemOpen
  \bibfield  {author} {\bibinfo {author} {\bibnamefont {{Aalbers, J. and
  others}}} (\bibinfo {collaboration} {{LUX-ZEPLIN}}),\ }\bibfield  {title}
  {\bibinfo {title} {{Background determination for the LUX-ZEPLIN dark matter
  experiment}},\ }\href {https://doi.org/10.1103/PhysRevD.108.012010}
  {\bibfield  {journal} {\bibinfo  {journal} {Phys. Rev. D}\ }\textbf {\bibinfo
  {volume} {108}},\ \bibinfo {pages} {012010} (\bibinfo {year}
  {2023})}\BibitemShut {NoStop}%
\bibitem [{\citenamefont {{P. Novella, B. Palmeiro, et
  al.}}(2018)}]{next_radon}%
  \BibitemOpen
  \bibfield  {author} {\bibinfo {author} {\bibnamefont {{P. Novella, B.
  Palmeiro, et al.}}} (\bibinfo {collaboration} {{NEXT}}),\ }\bibfield  {title}
  {\bibinfo {title} {{Measurement of radon-induced backgrounds in the NEXT
  double beta decay experiment}},\ }\href
  {https://doi.org/10.1007/JHEP10(2018)112} {\bibfield  {journal} {\bibinfo
  {journal} {J. High Energ. Phys.}\ }\textbf {\bibinfo {volume} {2018}},\
  \bibinfo {pages} {112}}\BibitemShut {NoStop}%
\bibitem [{\citenamefont {Agnes}\ \emph {et~al.}(2019)\citenamefont {Agnes}
  \emph {et~al.}}]{darkside50_ionfrac}%
  \BibitemOpen
  \bibfield  {author} {\bibinfo {author} {\bibfnamefont {P.}~\bibnamefont
  {Agnes}} \emph {et~al.} (\bibinfo {collaboration} {DarkSide}),\ }\bibfield
  {title} {\bibinfo {title} {{Measurement of the ion fraction and mobility of
  218Po produced in 222Rn decays in liquid argon}},\ }\href
  {https://doi.org/10.1088/1748-0221/14/11/P11018} {\bibfield  {journal}
  {\bibinfo  {journal} {{J. Instrum.}}\ }\textbf {\bibinfo {volume} {14}},\
  \bibinfo {pages} {P11018} (\bibinfo {year} {2019})}\BibitemShut {NoStop}%
\bibitem [{\citenamefont {{Albert and others}}(2015)}]{exo200_ionfrac}%
  \BibitemOpen
  \bibfield  {author} {\bibinfo {author} {\bibnamefont {{Albert and others}}}
  (\bibinfo {collaboration} {{EXO-200 Collaboration}}),\ }\bibfield  {title}
  {\bibinfo {title} {{Measurements of the ion fraction and mobility of
  $\ensuremath{\alpha}\text{\ensuremath{-}}$ and $\ensuremath{\beta}$-decay
  products in liquid xenon using the EXO-200 detector}},\ }\href
  {https://doi.org/10.1103/PhysRevC.92.045504} {\bibfield  {journal} {\bibinfo
  {journal} {{Phys. Rev. C}}\ }\textbf {\bibinfo {volume} {92}},\ \bibinfo
  {pages} {045504} (\bibinfo {year} {2015})}\BibitemShut {NoStop}%
\bibitem [{\citenamefont {Abi}\ \emph {et~al.}(2020)\citenamefont {Abi} \emph
  {et~al.}}]{DUNE:2020txw}%
  \BibitemOpen
  \bibfield  {author} {\bibinfo {author} {\bibfnamefont {B.}~\bibnamefont
  {Abi}} \emph {et~al.} (\bibinfo {collaboration} {DUNE}),\ }\bibfield  {title}
  {\bibinfo {title} {{Deep Underground Neutrino Experiment (DUNE), Far Detector
  Technical Design Report, Volume IV: Far Detector Single-phase Technology}},\
  }\href {https://doi.org/10.1088/1748-0221/15/08/T08010} {\bibfield  {journal}
  {\bibinfo  {journal} {{J. Instrum.}}\ }\textbf {\bibinfo {volume} {15}},\
  \bibinfo {pages} {T08010} (\bibinfo {year} {2020})}\BibitemShut {NoStop}%
\bibitem [{\citenamefont {{M. Adamowski and others}}(2023)}]{dune_cdr_2023}%
  \BibitemOpen
  \bibfield  {author} {\bibinfo {author} {\bibnamefont {{M. Adamowski and
  others}}} (\bibinfo {collaboration} {{LBNF/DUNE}}),\ }\bibfield  {title}
  {\bibinfo {title} {{LBNF/DUNE Cryostats and Cryogenics Infrastructure for the
  DUNE Far Detector, Design Report}},\ }\href@noop {} {\  (\bibinfo {year}
  {2023})},\ \Eprint {https://arxiv.org/abs/2312.09104} {arXiv:2312.09104
  [physics.ins-det]} \BibitemShut {NoStop}%
\bibitem [{\citenamefont {Machado}\ \emph {et~al.}(2019)\citenamefont
  {Machado}, \citenamefont {Palamara},\ and\ \citenamefont
  {Schmitz}}]{sbnd_phys}%
  \BibitemOpen
  \bibfield  {author} {\bibinfo {author} {\bibfnamefont {P.~A.}\ \bibnamefont
  {Machado}}, \bibinfo {author} {\bibfnamefont {O.}~\bibnamefont {Palamara}},\
  and\ \bibinfo {author} {\bibfnamefont {D.~W.}\ \bibnamefont {Schmitz}},\
  }\bibfield  {title} {\bibinfo {title} {{The Short-Baseline Neutrino Program
  at Fermilab}},\ }\href {https://doi.org/10.1146/annurev-nucl-101917-020949}
  {\bibfield  {journal} {\bibinfo  {journal} {Annu. Rev. Nucl. Part. Sci.}\
  }\textbf {\bibinfo {volume} {69}},\ \bibinfo {pages} {363} (\bibinfo {year}
  {2019})}\BibitemShut {NoStop}%
\end{thebibliography}%

\end{document}